\documentclass{nature3}
\usepackage{graphicx}
\usepackage{xcolor}
\usepackage{float}
\usepackage{siunitx}
\usepackage{longtable}
\usepackage{tabularx}
\usepackage{url}
\usepackage{amsmath}
\usepackage{amssymb}
\usepackage{subcaption}
\usepackage{courier}
\usepackage{adjustbox}
\usepackage{rotating}
\usepackage{inputenc}
\usepackage{textgreek}
\usepackage{booktabs}
\usepackage[labelfont=bf]{caption}
\usepackage{xcolor}
\usepackage{xspace}
\usepackage{threeparttable} 
\usepackage{booktabs}       

\usepackage[numbers]{natbib}

\usepackage[separate-uncertainty = true,multi-part-units=single, group-digits = integer, group-four-digits = true,per-mode=reciprocal]{siunitx}

\usepackage{astjnlabbrev}
\usepackage[colorlinks=true,citecolor=blue,urlcolor=cyan]{hyperref}
\usepackage[normalem]{ulem}

\usepackage[affil-it]{authblk}

\newcommand{\kms}{\mbox{${\rm km\,s}^{-1}$}}

\newcommand{\Mjup}{\mbox{${\rm M}_\mathrm{J}$}}

\newcommand{\Rjup}{\mbox{${\rm R}_\mathrm{J}$}}

\newcommand{\halpha}{\mbox{$H_\alpha$}}

\newcommand{\target}{WASP-121~b\xspace}
\newcommand{\targetsys}{WASP-121\xspace}

\usepackage[normalem]{ulem}
\usepackage[markup=bfit]{changes}
\definechangesauthor[name={Bibiana Prinoth}, color=purple]{BP}
\definechangesauthor[name={Julia Seidel}, color=violet]{JVS}

\DeclareRobustCommand{\ion}[2]{\textup{#1\,\textsc{\lowercase{#2}}}}
\makeatletter
\def\@maketitle{%
  \newpage
  \begin{center}%
  \let \footnote \thanks
    {\Large \@title \par}%
    \vskip -3em
    {\large
      \begin{tabular}[t]{c}%
        \@author
      \end{tabular}\par}%
  \end{center}%
  \par
  \vskip 0.5em}
\makeatother

\title{Vertical structure of an exoplanet's atmospheric jet stream}
\author[1,2,*, **]{Julia\,V.\ Seidel}
\affil[1]{European Southern Observatory, Santiago, Chile}
\affil[2]{Laboratoire Lagrange, Observatoire de la Côte d’Azur, CNRS, Université Côte d’Azur, Nice, France}

\author[1,3]{Bibiana Prinoth}
\affil[3]{Lund Observatory, Division of Astrophysics, Department of Physics, Lund University, Lund, Sweden}

\author[4]{Lorenzo Pino}
\affil[4]{INAF—Osservatorio Astrofisico di Arcetri, Florence, Italy}

\author[5,17]{Leonardo A. dos Santos}
\affil[5]{Space Telescope Science Institute, Baltimore, USA}
\affil[17]{William H. Miller III Department of Physics and Astronomy, Johns Hopkins University, Baltimore, USA}

\author[6]{Hritam Chakraborty}
\affil[6]{Observatoire de Genève, Département d’Astronomie, Université de Genève, Versoix, Switzerland}

\author[2]{Vivien Parmentier}

\author[1]{Elyar Sedaghati}

\author[8]{Joost P. Wardenier}
\affil[8]{Département de Physique, Institut Trottier de Recherche sur les Exoplanètes, Université de Montréal, Montréal, Canada}

\author[6]{Casper Farret Jentink}

\author[7]{Maria Rosa Zapatero Osorio}
\affil[7]{Centro de Astrobiología, CSIC-INTA, Madrid, Spain}

\author[8]{Romain Allart}
\affil[8]{Département de Physique, Institut Trottier de Recherche sur les Exoplanètes, Université de Montréal, Montréal, Canada}

\author[6]{David Ehrenreich}

\author[6]{Monika Lendl}

\author[9,10]{Giulia Roccetti}
\affil[9]{European Southern Observatory, Garching bei München, Germany}
\affil[10]{Meteorologisches Institut, Ludwig-Maximilians-Universität München, Munich, Germany} 

\author[11,12,1]{Yuri Damasceno}
\affil[11]{Instituto de Astrofísica e Ci\^encias do Espa\c{c}o, Universidade do Porto, CAUP, Porto, Portugal}
\affil[12]{Departamento de F\'isica e Astronomia, Faculdade de Ci\^encias, Universidade do Porto, Porto, Portugal } 

\author[6]{Vincent Bourrier}

\author[14]{Jorge Lillo-Box} 
\affil[14]{Centro de Astrobiolog\'ia (CAB), CSIC-INTA, Madrid, Spain}

\author[3]{H. Jens Hoeijmakers}

\author[15,16]{Enric Pallé}
\affil[15]{Instituto de Astrof\'{\i}sica de Canarias, La Laguna, Tenerife, Spain}
\affil[16]{Departamento de Astrof\'{\i}sica, Universidad de La Laguna, La Laguna, Tenerife, Spain}

\author[11,12]{Nuno Santos}

\author[15,16]{Alejandro Su{\'a}rez Mascare{\~n}o}
\affil[15]{Instituto de Astrof\'{\i}sica de Canarias, La Laguna, Tenerife, Spain}
\affil[16]{Departamento de Astrof\'{\i}sica, Universidad de La Laguna, La Laguna, Tenerife, Spain}

\author[11]{Sergio G. Sousa} 

\author[13]{Hugo M. Tabernero}
\affil[13]{Departamento de F{\'i}sica de la Tierra y Astrof{\'i}sica \& IPARCOS-UCM (Instituto de F\'{i}sica de Part\'{i}culas y del Cosmos de la UCM). Universidad Complutense de Madrid, Madrid, Spain}

\author[6]{Francesco A. Pepe} %

\affil[*]{Corresponding author: jseidel@oca.eu}
\affil[**]{ESO Fellow}

\begin{document}
\maketitle

\begin{abstract}

Ultra-hot Jupiters, an extreme class of planets not found in our solar system, provide a unique window into atmospheric processes. The extreme temperature contrasts between their day- and night-sides pose a fundamental climate puzzle: how is energy distributed?
To address this, we must observe the 3D structure of these atmospheres, particularly their vertical circulation patterns, which can serve as a testbed for advanced Global Circulation Models (GCM) \citep[e.g.][]{lee_mantis_2022}.
Here we show a dramatic shift in atmospheric circulation in an ultra-hot Jupiter: a unilateral flow  from the hot star-facing side to the cooler space-facing side of the planet sits below an equatorial super-rotational jet stream. 
By resolving the vertical structure of atmospheric dynamics, we move beyond integrated global snapshots of the atmosphere, enabling more accurate identification of flow patterns and allowing for a more nuanced comparison to models.
Global circulation models based on first principles struggle to replicate the observed circulation pattern \citep{ showman_equatorial_2011} underscoring a critical gap between theoretical understanding of atmospheric flows and observational evidence.
 This work serves as a testbed to develop more comprehensive models applicable beyond our Solar System as we prepare for the next generation of giant telescopes.

\end{abstract}

\subsection{WASP-121~b observations} 
\leavevmode\\
We present the first full transit dataset of an exoplanet observed with the four Unit Telescope (4-UT) mode of the Echelle Spectrograph for Rocky Exoplanets and Stable Spectroscopic Observations (ESPRESSO\citep{pepe_espresso_2021}) at ESO's VLT (European Southern Observatory's Very Large Telescope) located on Cerro Paranal, Chile. ESPRESSO is an ultra-stabilised, fibre-fed, high resolution spectrograph which can receive light from any or all of the four VLT 8-metre UTs. The 4-UT mode combines all four UTs to supply the spectrograph with the equivalent photon collecting power of a 16-metre class telescope. The 4-UT mode provides mid-resolution data ($\lambda/\Delta \lambda$ = 70,000) in 4$\times$2 binning. We combine archival data from the commissioning of the mode taken by the ESPRESSO Consortium covering the second half of the transit of the ultra-hot Jupiter WASP-121~b (taken on 30 November 2018, from here on called the egress epoch, \citet{ borsa_atmospheric_2021,  seidel_detection_2023-1}) with newly obtained data to complete the full transit (PI: Seidel, 23 September 2023, from here on called the ingress epoch). For the observation log for both epochs see Extended Data Fig. \ref{fig:transitOverview}. WASP-121~b is a canonical ultra-hot Jupiter orbiting its F6-type host star WASP-121 (V$_\mathrm{mag} = 10.4$) and is $16\%$ more massive than Jupiter while being $75\%$ larger in size ($1.74 \Rjup$, $1.157 \Mjup$, see Extended Data Table \ref{tab:parameters_W121}), making the planet highly inflated and about 2.5 times less dense than Saturn.
From the extremely high signal-to-noise ratio data of \target{}, we present the first vertical characterization of a high-altitude, super-rotational atmospheric jet stream. We also highlight its phase-resolved properties, showcasing changes in temperature and velocity as the jet transitions from the cold night-side to the day-side and later crosses back from the hot day-side to the night-side at the end of the transit.\\
\subsection{Dynamics explored across vertical layers} 
\leavevmode\\
\noindent We study three elements probing differing altitude levels to obtain the vertical wind structure of the atmosphere: \ion{Fe}{I} (deeper atmosphere), the \ion{Na}{I} doublet (mid-to-shallow atmosphere), and the $\halpha{}$-line of the Hydrogen Balmer-series (shallow atmosphere). We have opted for these three tracers due to their accessibility in ultra-hot Jupiter atmospheres and their complementary probing depth in pressure. 
Atmospheric dynamics in exoplanets are traced by the Doppler-shift induced on the spectral lines by the combination of atmospheric winds and planetary rotation in the planetary rest frame. With high S/N data such as in this work, this Doppler-shift can be measured as a function of orbital phase and thus give an evolution in time - the so called Doppler-trace (but see also work on smaller telescope on resolving the iron trace, e.g. \citet{ ehrenreich_nightside_2020, kesseli_confirmation_2021} and most recently the usage of iron lines to create a relative pressure profile from this species in \citet{ kesseli_up_2024}).

While we cannot provide true absolute pressure values due to the relative nature of echelle spectrograph observations, we can provide relative altitudes between the tracers for the cloud-free parts of the atmosphere assuming solar abundances and proper treatment of the largest scattering contributors to the continuum \citep{ Welbanks2019}.
\subsection{Deep atmosphere}
To access the circulation in the deepest layers of the atmosphere at the highest pressures, we use the cross-correlation method for \ion{Fe}{I}, which combines the plethora of shallow iron absorption lines in the visible wavelength range to produce an atmospheric Doppler trace of the atmosphere at relatively deep pressures (see left panel of Figure \ref{fig:altitude_profile}) \citep{ ehrenreich_nightside_2020, kesseli_confirmation_2021}. However, until now, the trace was created from stellar spectral line masks taking into account all species in the stellar atmosphere which is inaccurate for spectra not dominated by iron (see Methods \ref{sec:planetaryCCF} for an in-depth discussion and further detail on the creation of Doppler-shift traces from cross-correlation). Here, we provide the atmospheric trace with a line list generated solely from \ion{Fe}{I} probing deeper pressures (lower altitudes) in the atmosphere \citep[from][]{ kitzmann_mantis_2023}. The contribution function which shows the impact of each pressure layer on the cross-correlation result was generated using the radiative transfer code \texttt{shone} taking into account scattering (H$-$ and Rayleigh) to generate an absolute pressure range.
\subsection{Mid atmosphere}
Complementary, resolved planetary spectral lines probe larger pressure ranges at shallower absolute pressures at the order of magnitude level (higher in altitude). The resolved \ion{Na}{I} Fraunhofer doublet probes roughly below the temperature inversions which canonically indicate the thermosphere \citep{ wyttenbach_spectrally_2015,  huang_model_2017}. From the contribution function of our \ion{Na}{I} model (calculated with MERC, see methods \ref{sec:contribution_func}) we find that the bulk of the features are generated above the probing range of \ion{Fe}{I} (see middle panel of Figure \ref{fig:altitude_profile}). We cross-validated between our tracers by also generating the cross-correlation for \ion{Na}{I} lines, arriving at the same continuum as with the resolved spectral line model. MERC creates a quasi-3D atmospheric grid from an isothermal temperature profile and overlays planetary rotation with common atmospheric wind patterns in a lower and upper layer, such as sub-to-anti-stellar-point flow (a unilateral flow across the planet from the hot to cold side), super-rotational, and radial winds mimicking atmospheric mass loss. These patterns are then translated into the combined impact of their line-of-sight components on the spectral lines and compared to the obtained data via a Bayesian retrieval framework which also retrieves the continuum level and the temperature profile (for more details see Methods \ref{sec:merc}).
\subsection{Shallow atmosphere}
To supplement the vertical probing range of \ion{Na}{I}, we employ the \halpha{}-line of the Balmer-series. The contribution of the atmosphere for deeper pressures than the \si{\micro\bar} level to the \halpha{}-line is negligible \citep{ huang_hydrodynamic_2023} and thus provides a complimentary probe to \ion{Na}{I} for shallower pressures (higher altitudes). The contribution function (right panel Figure \ref{fig:altitude_profile}, generated with {\tt p-winds}, see Methods \ref{sec:pwinds}) confirms this finding and highlights that, assuming the same system parameters and proper treatment of scattering processes, the \halpha{}-line has a non-negligible overlap with the deeper pressure ranges of the \ion{Na}{I} doublet. The {\tt p-winds} implementation of Parker-wind outflow for the Balmer-series as well as priors implemented for WASP-121~b are described in Methods \ref{sec:pwinds}.
\subsection{Dynamics explored horizontally} 
\leavevmode\\
The exceptional S/N in a single transit with ESPRESSO's 4-UT mode allows to phase-resolve all chosen atmospheric tracers. Phase-resolving tracers subsequently permits to draw conclusions from the changing viewing geometry, e.g. at the beginning of the transit, the probed atmosphere is dominated by the leading limb (planetary morning) and a portion of the hot day-side of the planet while the trailing limb (planetary evening) is dominated by the cooler night side. This geometry then reverses at the end of the transit (see Figure \ref{fig:drawing} for a reference).
For the target WASP-121~b, the angle of the planet's terminator compared to the line of sight (located at 0$^\circ$ at mid-transit, or orthogonal to the line of sight and marked in Figure \ref{fig:drawing} with $\theta$) changes by nearly 20$^\circ$ between transit start and mid-transit, with ingress, where the planet is only partially in front of the star (phases $-0.0475$ to $-0.0372$) spanning viewing angles from $-17^\circ$ to $-13.5^\circ$ and egress vice versa. Phase-resolving more prominently highlights the effects related to the dependence of the spectrum on the position on the stellar limb as they are not averaged out across the transit cord. For the specific case of WASP-121~b, the largest residual impact of the RM-effect occurs during mid-transit (see Extended Data Figure \ref{fig:tracks} for a view of the overlap between the stellar and planetary signal). We have in consequence opted to discard the affected phases conservatively between $-0.017$ and $0.017$ and only draw conclusions for the phases before and after the exclusion zone where any stellar impact is far removed in velocity space from the planetary atmospheric features due to the relative motion of the planet. 
\subsection{Probing the perpetual planetary morning and evening}
This leaves us with two sets of data with ten exposures each, one from the ingress epoch observations taken in 2023, covering a viewing angle from $-17^\circ$ to $-7^\circ$ (from here called the morning segment, due to its dominant coverage of the perpetual planetary morning on the leading limb of the planet) and one from the archival egress epoch published in \citet{ borsa_atmospheric_2021} covering a viewing angle from $7^\circ$ to $17^\circ$ (from here called the evening segment, due to its dominant coverage of the planetary evening on the trailing limb of the planet). The mean geometry and terminology is shown in Figure \ref{fig:drawing}. For each segment we binned the corresponding data to achieve the necessary signal to noise ratio to resolve the spectral lines. For \ion{Na}{I} we combined all ten exposures of each segment (following the approach in \citet{ seidel_detection_2023-1} for the previous analysis of the evening segment), for the \halpha{}-line we continuously combined four exposures for the entire transit (see Figure \ref{fig:Halpha_resolved}) but excluded data with RM-effect overlap from our conclusions (see also Extended Data Table \ref{tab:halpha_comparison}). Lastly, no binning was necessary for the cross-correlation analysis.\\
\subsection{Observed atmospheric dynamics} 
\leavevmode\\
Probing deepest in the atmosphere is the atmospheric \ion{Fe}{I} trace (shown in green in Figure \ref{fig:drawing} and in Figure \ref{fig:doppler_overview}). For the \ion{Fe}{I} trace, we find line-of-sight blueshifts both for the morning and evening segment ($-4.12$ and $-6.90~\kms{}$ respectively), indicating a sub-to-anti-stellar-point flow ranging from approximately $-6$ to $-10~\kms{}$ (this de-projection transformation is based on the wind implementation and geometry of MERC\citep{ seidel_wind_2020,  seidel_into_2021}). \\
\noindent The combined lines of the \ion{Na}{I} doublet probe ranges above the \ion{Fe}{I} trace and are unaffected by the sub-to-anti-stellar-point wind probed by \ion{Fe}{I} (see Figure \ref{fig:altitude_profile}, for a comparison of their pressure profiles). 
\subsection{\ion{Na}{I} probing the super-rotational jet stream}
For the evening segment, where iron moves towards the observer in the deeper atmosphere, \citet{ seidel_detection_2023-1} detected a highly blue-shifted secondary \ion{Na}{I} peak offset from the line centre at zero velocity (see right panel of Figure \ref{fig:sodium_model_overview} and Figure \ref{fig:doppler_overview}) and employed the MERC code to deduce the atmospheric wind patterns at the root of the secondary peak. They hypothesized that the separated \ion{Na}{I} feature could be due to either a super-rotational jet stream or a sub-to-anti-stellar-point wind, as seen for iron, both constrained to latitudes away from the poles ($\sim 30^\circ$ from the equator for a jet opening angle of $\sim 60^\circ$ as an order of magnitude estimation - less than half the latitude range). Due to the lack of data during the first third of the transit which provides information on the planet's morning, they were unable to discriminate between the two scenarios. We repeat their results in Extended Data Table \ref{table:sodium_overview} and complete their preliminary analysis with the morning segment.

We find another secondary atmospheric \ion{Na}{I} feature for the morning segment, however, contrary to the evening segment or the iron trace the feature is red-shifted compared to the central \ion{Na}{I} wavelength indicating movement away from the observer which is fundamentally incompatible with the imprinted movement towards the observer from a sub-to-anti-stellar-point flow (see Figure \ref{fig:drawing}). The forward models retrieved on the morning segment with a super-rotational atmosphere are preferred over the dynamically quiet model with odds of 13:1, while constraining the jet to less than half the latitude range latitude of the equator as in \citet{ seidel_detection_2023-1} leads to odds of 56:1 over the quiet model, odds of 42:1 over the wind models without super-rotation, and 4:1 over the super-rotational model with no latitude constraint. The Bayesian evidence comparison for this analysis of the morning segment is discussed in Methods \ref{sec:merc}, and shown in Extended Data Table \ref{table:comparison_model}.

\noindent In conclusion, the most likely scenario combining the knowledge from the planetary evening and morning segments is a sub-to-anti-stellar-point flow traced by iron in the deep atmosphere sitting below a super-rotational movement crossing from the morning terminator to the evening terminator across the day-side at higher altitudes, likely constrained to less than half the latitude range from the equator. This leads to a red-shifted feature during the morning segment probing the leading limb pointing away from the observer and a blue-shifted feature during the evening segment probing the trailing limb (see Figure \ref{fig:drawing} for a visual of the geometry and the observed atmospheric dynamics).
We find a significant temperature increase of $950\pm560$~K between the morning segment and the evening segment indicating that the material in the jet stream is heated as it crosses the day-side of the planet, however, the location of a potential hotspot cannot be constrained and is likely variable \citep{ changeat_is_2024}. Additionally, the wind speed of the observed jet-stream increases from $13.7\pm6.1\kms{}$ in the morning segment to $26.8\pm7.13\kms{}$ in the evening segment as retrieved with MERC, see Methods \ref{sec:merc} indicating a thus far never observed acceleration as the jet traverses the day-side. Curiously, while the jet-stream fit is clearly accompanied by a vertically outwards pointing wind above the jet-stream boundary in the evening segment probing planetary evening \citep{ seidel_detection_2023-1}, the wind speed of this vertical wind was only tentatively constrained for the morning segment probing planetary morning. Combined with the temperature increase from the morning segment to the evening segment, it is likely that \ion{Na}{I} probes mostly the super-rotational jet stream and less the material outflow at higher altitudes during the morning segment. When we reach the evening terminator probed mostly during the evening segment, the higher temperatures translate to an increased scale height and in consequence \ion{Na}{I} additionally probes the beginning of material outflow, overlapping with the probing range of $\halpha$ (see Figure \ref{fig:drawing} where \ion{Na}{I} is indicated in yellow for a visual). While the viewing geometry only allows us to probe the jet at its very start and end, tentative detections of strong velocity offsets for \ion{Na}{I} were reported for WASP-121~b via emission spectroscopy probing the central day-side which were not seen for lower probing species \citep{ hoeijmakers_mantis_2024} providing further evidence for the high altitude, super-rotational jet stream.\\
\subsection{Vertical extension to the shallow atmosphere}
\noindent To validate the jet stream observation, we employ the $\halpha{}$-line which starts probing the atmosphere at pressure ranges which should in theory still be impacted by the upper portion of the super-rotational jet stream (see Figure \ref{fig:altitude_profile}). The spectra stacked as a function of phase are shown in Figure \ref{fig:Halpha_resolved} and highlighted in blue together with the traces of \ion{Na}{I} and \ion{Fe}{I} in Figure \ref{fig:doppler_overview}. Taking into account planetary rotation, we retrieve an overall red-shift for \halpha{} translating to zonal winds speeds of $5.2\pm1.4\kms{}$ for planetary morning which then slowly turns to a blue-shift ($-19.2\pm1.4\kms{}$) as planetary evening becomes dominant. This indicates the same super-rotational movement observed in the jet stream for \ion{Na}{I} imprints on the outwards flow of hydrogen which is usually a solely vertical Parker-wind motion. Combined with the tentative detection of a vertical outflow above the jet stream for \ion{Na}{I} this shows that the probing ranges of \ion{Na}{I} and $\halpha{}$ are not mutually exclusive but validate each others findings (see also Figure \ref{fig:altitude_profile} for a quantitative assessment). Summarising the observed atmospheric dynamics, we find a day-to-night side movement traced by \ion{Fe}{I} below an equatorial jet stream, a vertical profile thus far never observed in an exoplanet atmosphere. This shift from day-to-night winds to the jet steam might first be seen with the full atmospheric trace, while the bulk of the jet stream is probed via the \ion{Na}{I} doublet. Due to the overlap in pressure ranges between the \ion{Na}{I} and the $\halpha$-line both wind patterns likely impact $\halpha$ - the radial, vertical flow via Parker-winds which adds symmetrical broadening as well as the jet stream which induces line centre Doppler-shifts on the $\halpha$-line. The vertical winds associated with Parker-winds additionally also tentatively imprint on the \ion{Na}{I} doublet in the hottest probed parts of the atmosphere providing two-way evidence of the shift in the vertical profile. \\  
\subsection{Comparison with theoretical models} 
\leavevmode\\
Comparing our observational results with theoretical predictions from \citet{ lee_mantis_2022}, global circulation models (GCMs) conducted at the \SI{}{\micro \bar} pressure level for WASP-121~b predict a jet-stream constrained to $\sim 30-40^\circ$ in latitude ranging from the deeper atmosphere to at least the end of their pressure grid at $0.1$\SI{}{\micro \bar}. In said work, they additionally predict a temperature gradient at these altitudes with a temperature increase from the morning terminator to the evening terminator by $\sim 1000~K$. While this prediction is confirmed by our observations (under the caveat that the pressure scales between these works align sufficiently and the GCM results at the end of their pressure grid can be extrapolated upwards), our observed jet-stream velocities lie above of the predictions in \citet{ lee_mantis_2022}. One additional factor which might lead to the stronger velocity profile could be the impact of magnetic fields as described in \citet{ beltz_magnetic_2023} and the impact of H$_2$ dissociation/recombination which is not included in the GCM from \citet{ lee_mantis_2022}. 
A comparison of the atmospheric \ion{Fe}{I} trail with GCMs from \citet{ wardenier_phase-resolving_2024} in Figure \ref{fig:atmospheric_shimmer_all} suggest that the observations are consistent with (at most) relatively weak uniform drag and additional heat transport due to hydrogen dissociation/recombination \citep{ Komacek2018,  Bell2018,  tan_modelling_2024}. Reducing the drag timescale or switching off hydrogen dissociation/recombination results in the model trails being significantly more redshifted than our data. However, as already discussed in \citet{ wardenier_phase-resolving_2024}, none of the four presented GCM trails provide a perfect match to the \ion{Fe}{i} trail (neither for the iron trace as derived from the stellar spectrum, closest represented by the atmospheric trace generated from all spectral lines as used in \citet{ wardenier_phase-resolving_2024}, nor for the more accurate \ion{Fe}{i} trail from an iron-only line list, see Figure \ref{fig:contributionTrace}). That is, in the evening segment, all models accounting for hydrogen dissociation/recombination overestimate the blueshift of the iron lines by a few km/s. While the exact cause for this discrepancy is unknown, it is important to stress that \citet{ wardenier_phase-resolving_2024} only tested four models, which is by no means a full exploration of the parameter space. Moreover, all models assume a uniform drag timescale across the atmosphere, which is a simplifying assumption made in the GCM. In reality, the drag timescale can vary by orders of magnitude between the day-side and the night-side of the planet\citep{ beltz_magnetic_2023}, leading to a different circulation profile. Follow-up studies will need to shed light on the additional physics, such as possible additional XUV energy deposit from the host star on the day-side, that needs to be included in GCMs to provide a better match to the extreme cases of ultra-hot Jupiters. 

The discrepancy between GCMs and the provided observations highlight the impact of high signal-to-noise ratio data of extreme worlds such as ultra-hot Jupiters in benchmarking our current understanding of atmospheric dynamics. This study marks a shift in our observational understanding of planetary atmospheres beyond our solar system. By probing the atmospheric winds in unprecedented precision, we unveil the 3D structure of atmospheric flows, most importantly the vertical transitions between layers from the deep sub-to-anti-stellar-point winds to a surprisingly pronounced equatorial jet stream. These benchmark observations made possible by ESPRESSO's 4-UT mode serve as a catalyst for the advancement of global circulation models across wider vertical pressure ranges thus significantly advancing our knowledge on atmospheric dynamics. As we ramp up for the future of exoplanet observations with Extremely Large Telescopes (like ESO's ELT), extremely high signal-to-noise ratio observations will become the gold standard to test whether our models capture sufficient physical detail to explain the atmospheres of far-away worlds.

\clearpage

 \begin{figure*}
   \centering
\resizebox{\columnwidth}{!}{\includegraphics[trim=0.0cm 0.0cm 0.0cm 0.0cm]{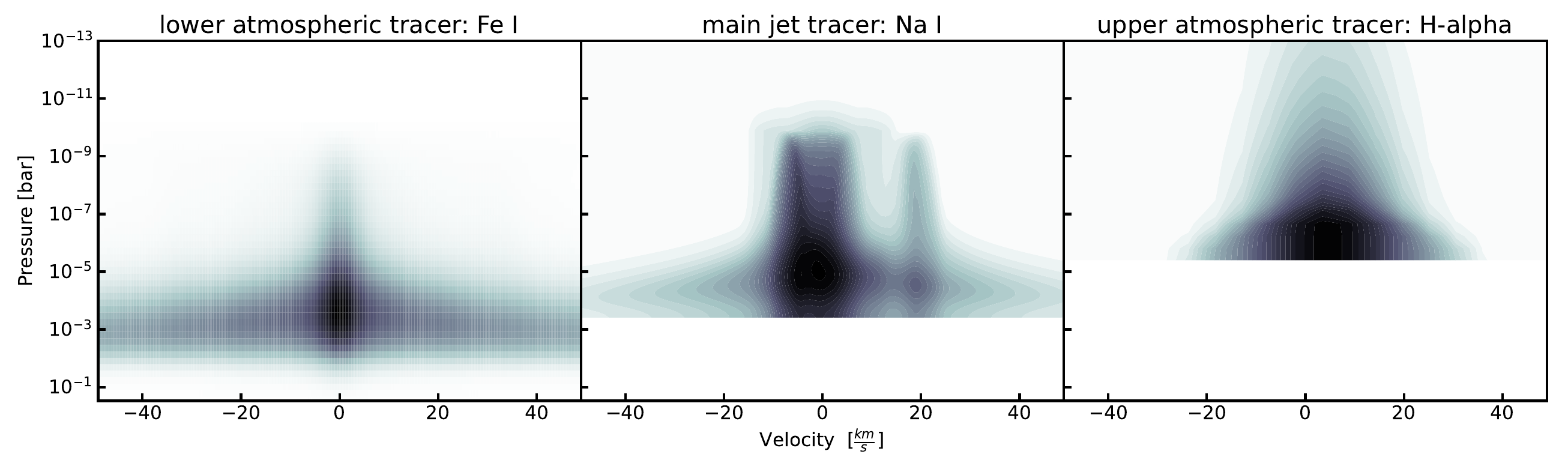}}
      \caption{Probed pressure ranges per tracer. From left to right the panels show the contribution function in for iron, sodium, and hydrogen (\halpha{}) (with lower pressures corresponding to higher altitudes) highlighting the overlapping altitude ranges. The base radius was set to the white light radius and the corresponding pressure retrieved accordingly. Left: Iron contribution function in cross-correlation with a large continuum contribution with little impact from the stronger lines. Middle: Contribution function for the morning segment from the \ion{Na}{I} MERC retrieval in velocity space showing the clear impact of the jet with a secondary feature at $20~\kms$. Right: Contribution function calculated from p-winds for the first exposure in $\halpha{}$. The colour contours indicate the strength of the contribution, with black as maximum contribution and white as no contribution.}
         \label{fig:altitude_profile}
   \end{figure*}

   \begin{figure*}[ht]
    \centering
   
    \begin{subfigure}{\textwidth}
        \centering
\resizebox{\columnwidth}{!}{\includegraphics[trim=2.0cm 9.0cm 2.0cm 0.0cm]{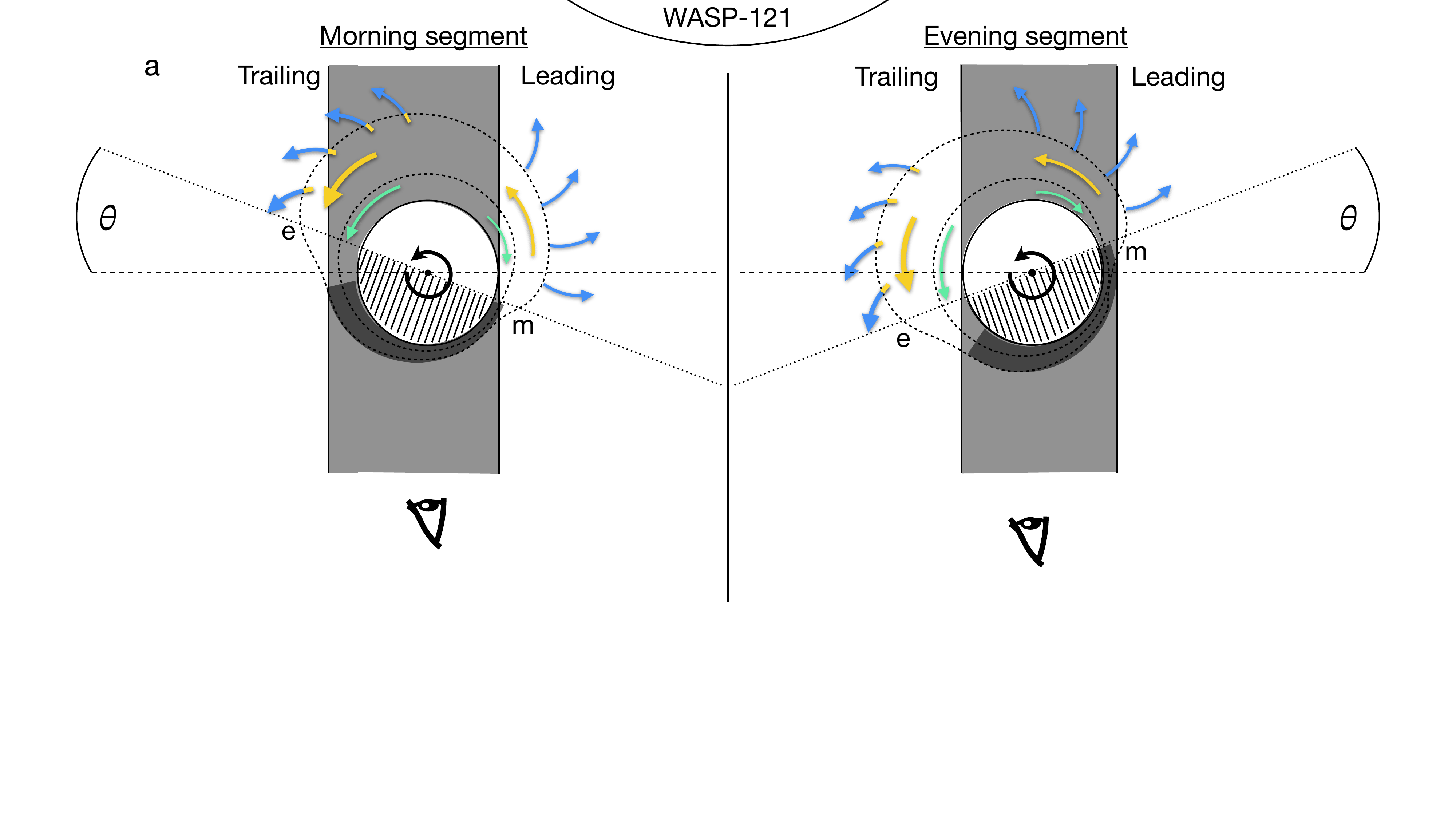}}
    \end{subfigure}
    \vspace{1em} 
    \begin{subfigure}{\textwidth}
        \centering
        \resizebox{\columnwidth}{!}{\includegraphics[trim=4.5cm 0.0cm 0.0cm 0.0cm]{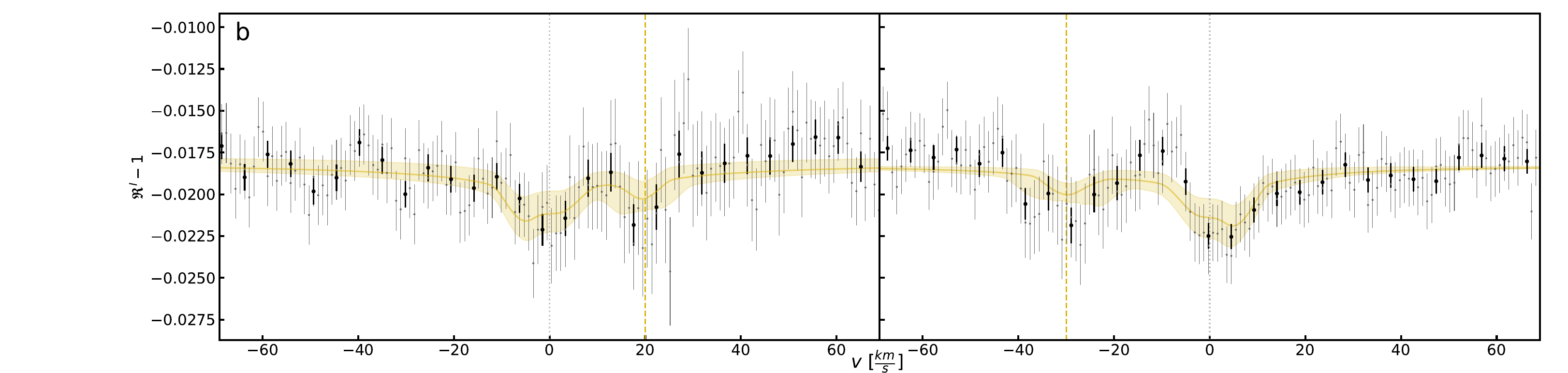}}
        \phantomsubcaption
         \caption*{Figure 2: The jet stream probed by sodium. (a) Diagram of the viewing geometry of WASP-121~b during the morning and evening segment (left and right) (not to scale) highlighting the three probing species and their movement direction over the planet's surface. The polar view is shown with the observer on the bottom and the host star on the top. The parts of the line of sight obscured by the planetary disk are shaded in grey, the night side of the planet is line-hatched. Dark grey indicates likely night-side cloud cover as modelled for WASP-121~b in \citep{helling_cloud_2021}, see Methods \ref{sec:merc} for details. The direction of planetary rotation (assumed from synchronous rotation with the star) is shown as a black arrow around the pole, planetary winds are shown as arrows with the colour indicating the probing element (light green: \ion{Fe}{I}, yellow: \ion{Na}{I}, dark blue: $\halpha$). The rough shape of the atmosphere as derived from \citet{wardenier_phase-resolving_2024} is shown as black dashed outlines. The trailing and leading limb are marked as well as the evening terminator (e) and the morning terminator (m) where the transition from permanent day- to night-side occurs and vice versa. The angle indicates the angle of the viewing geometry as described in the manuscript.}
                 \label{fig:drawing}
        \phantomsubcaption
        \caption*{(b) Evolution of the jet-induced secondary sodium feature from red-shifted to blue-shifted from the morning to the evening segment. The \ion{Na}{I} doublet is combined in velocity space for the morning segment (left) and evening segment (right). The data is shown in grey, and binned by 5 for visibility. The continuum is adjusted for the white light radius contribution of the planetary disk. The $1\sigma$ uncertainties have been propagated accordingly from the errors calculated by the ESPRESSO pipeline. In golden the best fit model from the MERC retrieval is shown with its $1\sigma$ uncertainty envelope (generated from the best fit values and uncertainties in \citet{seidel_detection_2023-1} for the evening segment).}
            \label{fig:sodium_model_overview}
    \end{subfigure}
\end{figure*}

    \begin{figure*}
   \centering
\resizebox{\columnwidth}{!}{\includegraphics[trim=0.0cm 0.0cm 0.0cm 0.0cm]{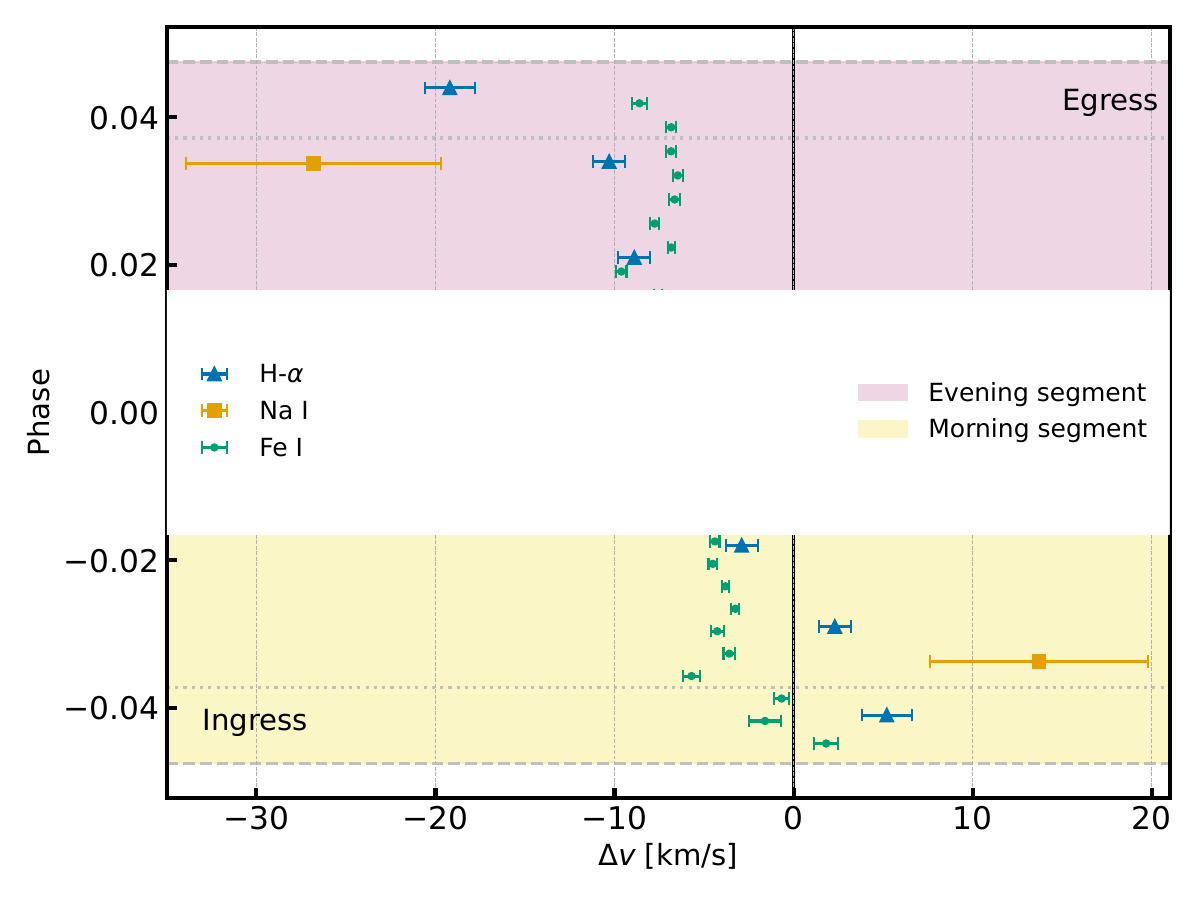}}
      \caption{Distinct movements for the three tracers. The atmospheric tracer shifts highlight the different movement pattern of the sub-to-anti-stellar-point wind and the super-rotational jet stream. The difference in movement patterns highlights the different regimes, with iron as green dots showing sub- to anti-stellar flow and sodium (yellow squares) as well as $\halpha{}$ (blue triangles) probing the jet with its emblematic red- to blue-shifted signature in time. The zero velocity, indicating a stationary atmosphere is highlighted as a dark vertical line. The morning and evening segment are highlighted with yellow and red shading, with the spectra taken with partial visibility of the atmosphere in ingress and egress separated by dotted lines. The data impacted by the RM-effect is omitted. The $1\sigma$ uncertainties have been propagated accordingly from the errors calculated by the ESPRESSO pipeline.}
         \label{fig:doppler_overview}
   \end{figure*}

   \begin{figure}[ht]
    \centering
    \begin{subfigure}[b]{0.49\textwidth}
       \resizebox{\columnwidth}{!}{\includegraphics[trim=3.0cm 0.0cm 1.0cm 0.0cm]{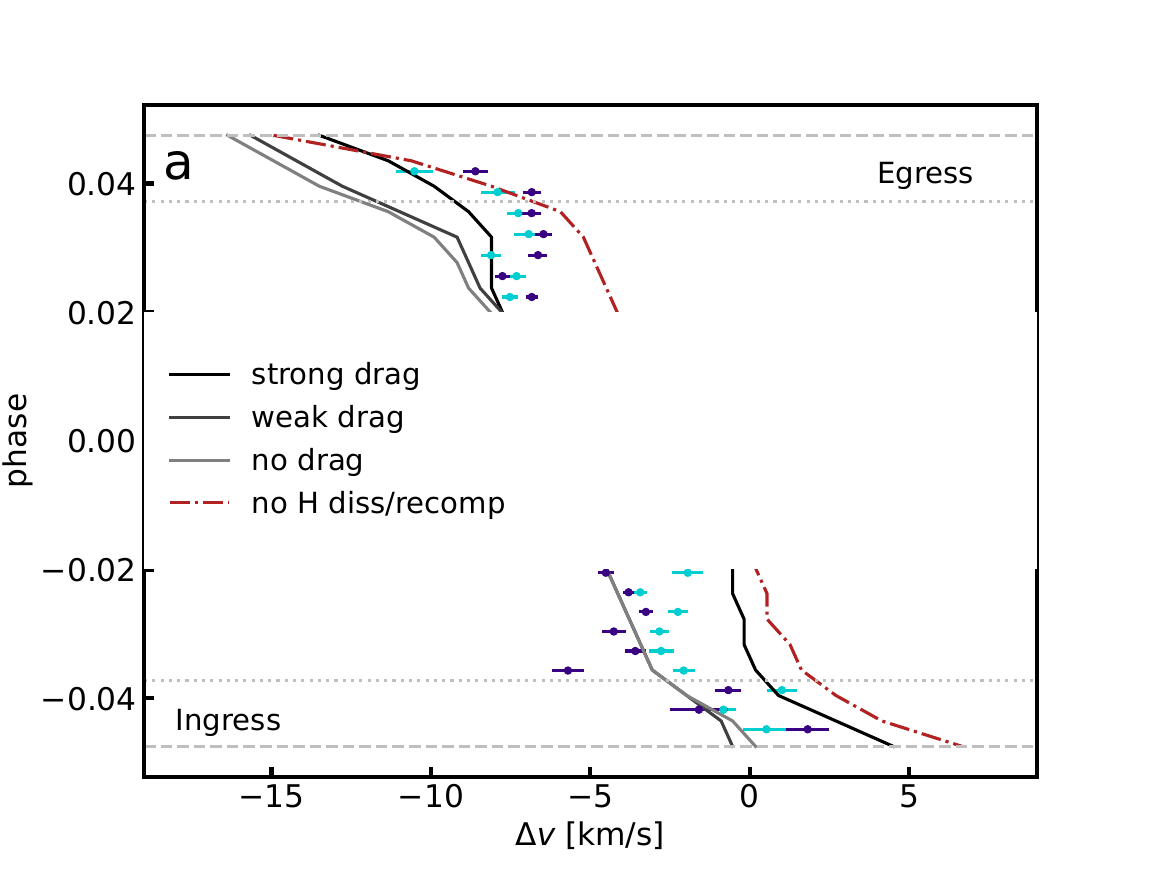}}
    \end{subfigure}
    \hfill
    \begin{subfigure}[b]{0.49\textwidth}
        \resizebox{\columnwidth}{!}{\includegraphics[trim=0.0cm 0.0cm 2.0cm 0.0cm]{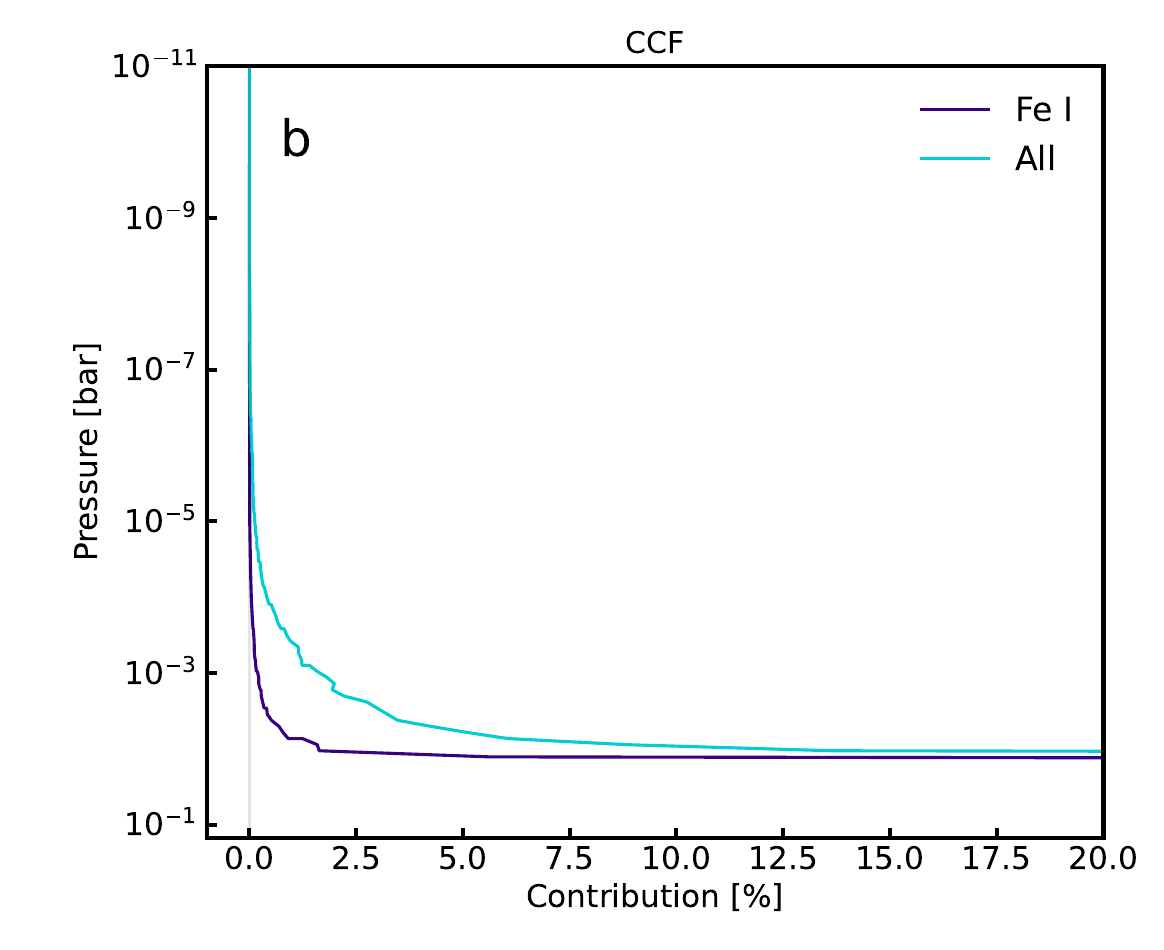}}
    \end{subfigure}
        \phantomsubcaption
          \label{fig:atmospheric_shimmer_all}
       \caption*{Drag as an insufficient explanation for the observed traces. Figure 4: (a) Favouring of no or weak drag models from \citet{ wardenier_phase-resolving_2024} for lower atmospheric tracers. We show two datasets: the \ion{Fe}{I} trace generated from a template containing solely iron lines and the all trace which includes all atmospheric species (All), mimicking that data as prepared for \citet{ wardenier_phase-resolving_2024} for better comparison. The centre of the in-transit \ion{Fe}{I} and All CCFs Gaussian fits are shown as a function of the orbital phase which corresponds to the line of sight velocity. The $1\sigma$ uncertainties have been propagated accordingly from the errors calculated by the ESPRESSO pipeline. Horizontal lines show the transit duration, dashed lines show the full-transit limits. All exposures impacted by the RM-effect were masked. The overplotted models are taken from \citet{ wardenier_phase-resolving_2024}.}
              \phantomsubcaption
    \caption*{Figure 4: (b) Inconsistency between low atmospheric tracers generated from an \ion{Fe}{I} template and from the all lines stellar template as employed in \citet{ borsa_atmospheric_2021,  wardenier_phase-resolving_2024}. The figure shows a comparison between \ion{Fe}{I} and All CCFs Gaussian fits with the contribution histogram as a function of pressure. It shows that the data used in \citet{ wardenier_phase-resolving_2024}, while dominated by iron, likely contains components probing at lower pressures, and higher altitudes, than a pure iron analysis.}
    \label{fig:contributionTrace}
\end{figure}

\begin{figure}
   \centering
\resizebox{\columnwidth}{!}{\includegraphics[trim=-3.0cm 1.0cm -3.0cm 0.0cm]{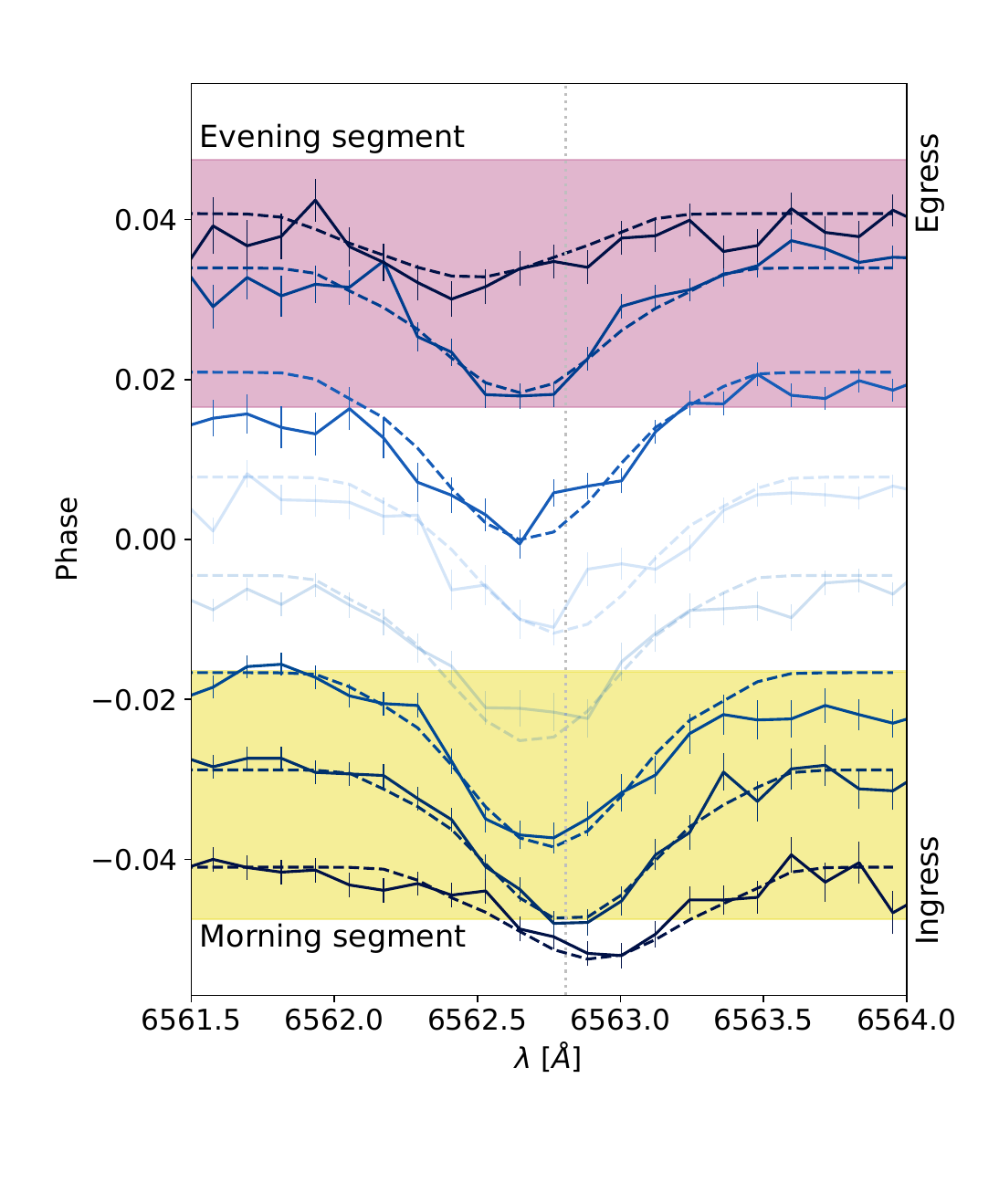}}
      \caption{Jet stream impacts high altitudes. Increasing blueshift of the $\halpha{}$ line indicating the impact of the jet steam even at high altitudes. The line centre shows a clear shift from redder wavelength towards a strong blueshift at the end of the evening segment. The data was binned by 4 exposures and 5x in wavelength. The $1\sigma$ uncertainties have been propagated accordingly from the errors calculated by the ESPRESSO pipeline. The flux values are scaled to the approximate phase values of the central exposure to highlight the evolution in time and the two datasets containing only ingress and egress marked. The vertical, dotted line indicates the central wavelength of $\halpha$. The dashed models for each phase correspond to the retrieved best fit model spectra from {\tt p-winds} with the mass loss rate taken from literature. While shown with reduced opacity, phases from $0.017$ to $-0.017$ are affected by the RM-effect and were excluded from the conclusions.}
         \label{fig:Halpha_resolved}
\end{figure}

\clearpage



\begin{methods}

\section{Resolved \ion{Na}{I} doublet and \halpha-line}
\label{sec:methods}
In this work, we explore the phase-resolved \ion{Na}{I} doublet and \halpha line via transmission spectroscopy. To dissentangle the planetary signal from the stellar spectral lines, we follow the established procedure to extract the planetary signal through its offset in velocity space due to the relative radial velocity between the star and the planet. For more details,  see e.g. \citet{seidel_detection_2023-1}. In \citet{seidel_detection_2023-1}, the egress epoch presented in \citet{borsa_atmospheric_2021} was used to detect the high-velocity \ion{Na}{I} feature. Subsequently, to pinpoint the location of said feature in phase (and thus location in longitude) the data was separated into subsets, one centred around mid-transit from orbital phases $-0.017$ (start of observations) to $0.017$, and one starting at $0.017$ until the end of egress at phase $0.048$. In this work, for consistency with \citet{seidel_detection_2023-1}, we separate the ingress epoch 4-UT data in two subsets, the morning segment ranging from ingress at $-0.048$ to phase $-0.017$ (a total of 10 exposures) and the centre segment from $-0.017$ until $0.009$ (end of observations, total of 9 exposures). The ESPRESSO ADC (atmospheric dispersion corrector) is unable to properly correct for atmospheric dispersion for airmasses above 2.2. As a consequence we have discarded the first two exposures of the ingress epoch. We do not re-analyse the egress epoch analysis of the \ion{Na}{I} doublet, but provide the results from \citet{seidel_detection_2023-1} for context. The analysis steps provided below are applied to the ingress epoch.

In transmission spectroscopy, the various spectral lines, telluric, stellar, planetary, and interstellar medium in some cases, will be Doppler-shifted by their respective velocities. As a consequence, different contamination sources, such as telluric spectral lines or the Rossiter-McLaughlin effect (RM), only impact the planetary signal during specific times of the orbit where they overlap. For the data analysed in this work, the impact of the different contamination sources with respect to the planetary trace are shown in velocity space in Figure \ref{fig:tracks} for both observing epochs. For both epochs, the centre segment is impacted by the residuals from stellar lines and the RM-effect. The morning segment, which is instrumental in our jet-stream detection (location marked in light blue for both epochs in Figure \ref{fig:tracks}) shows overlap with the telluric contaminants which could possibly provide a false-positive detection if not properly corrected for.

\section{Impact on the line shapes}
\label{sec:contaminants}
\subsection{Telluric correction}

We correct for telluric spectral lines generated in Earth's atmosphere via {\tt molecfit} (version 4.3.1 run on the ESOReflex environment, version 2.11.5) \citep{smette_molecfit_2015, kausch_molecfit_2015}. We applied {\tt molecfit} to correct for H$_2$O and O$_2$ primarily with the parameters set as in \citet{allart_search_2017} and refer to this work for further information. We find no remnants of tellurics or over-correction on any order of exposure three, which is the spectrum with the highest airmass. It should in theory be the most affected by telluric absorption features and its correction is shown in in the Supplementary Material. A visual inspection both of the individual spectra and the combined master spectrum showed no residuals from the telluric correction.

\subsection{Telluric emission}

While telluric spectral lines can be corrected for by modelling atmospheric conditions during the time of observation, other atmospheric effects are less predictable. Telluric emission plays an important role in the wavelength range of the \ion{Na}{I} doublet. The telluric \ion{Na}{I} layer in Earth's mesosphere can be excited and produces nightglow with seasonal variation strongest in winter \citep{herschbach_excitation_1992}. This emission feature is visible in the spectra as emission peaks at the location of the \ion{Na}{I} doublet in the observer's rest frame. While our observations were taken close to the equinox where the telluric \ion{Na}{I} layer plateaus, the layer can also be excited by external stimulus from meteor showers \citep{michaille_characterization_2001}. For transits, the emission background is routinely measured in parallel with a second fibre (fibre B) set on sky. For the ingress epoch, fibre B showed an emission \ion{Na}{I} doublet at near constant strength throughout the observations (see Extended Data Fig. \ref{fig:sodium_emission}). A possible source of the observed excitation of the atmospheric \ion{Na}{I} layer during our observations is the Daytime Sextantid meteor shower (221 DSX) in the Southern Hemisphere with its peak on 27-Sept-2023.\\

\noindent In theory, the sky-subtracted data as provided by the ESPRESSO pipeline should remove any sky features. However, as a consequence the overall noise is also increased and the quality of the subtraction cannot be verified. We have, therefore, opted to use the blaze corrected data products and have corrected manually for the emission feature. In Extended Data Figure \ref{fig:sodium_emission} a Gaussian fit to the doublet is shown. Given how narrow the emission feature is ESPRESSO's resolution is not sufficient to properly resolve the line peak and the fit does not capture the line shape at line centre. The \ion{Na}{I} emission crosses directly over the wavelength range of the jet feature (see Figure \ref{fig:tracks}) and any residual from a partial correction could alter the line shape. In consequence, we have opted to conservatively mask the affected wavelength region in the observer's rest frame. The masked windows are shown in Extended Data Figure \ref{fig:sodium_emission} as yellow background boxes and span the wavelength ranges from $5890.05$ to $5890.35~\AA$ and from $5896.03$ to $5896.33~\AA$ in the observer's rest frame, the maximum span of the emission feature in all exposures of the ingress epoch.\\
For the egress epoch, telluric contamination is discussed in \citet{seidel_detection_2023-1}. While fibre B was not set to sky during the egress epoch, the relative velocity offset between tellurics and the jet-feature is large enough to rule out contamination. Additionally, \citet{seidel_detection_2023-1} concluded that no meteor showers occured during the time period which would excite the sodium layer.

\subsection{Doppler-smearing}

\noindent While the various shifts from the observer's rest frame to the planetary rest frame can be corrected for, Doppler-smearing is intrinsic to the observing setup. It stems from the planetary movement along its orbit during an exposure and produces additional broadening of all spectral features \citep{wyttenbach_mass-loss_2020, ridden-harper_search_2016, cauley_2021}. The exposure time is mainly driven by the need for high S/N and Doppler-smearing cannot be avoided for fast moving planets such as ultra-hot Jupiters, however the trade-off should be considered carefully \citep[see e.g.][ for an in-depth discussion of this issue]{boldt-christmas_optimising_2024}. For the selected exposure time of 300s (to match the existing egress epoch) the maximum smear for each exposure of WASP-121~b is $1.1\, \kms$ corresponding to one resolution element in 4-UT MR mode. Because this smearing is smaller than the uncertainty on the data ($\sim2\kms$), the impact on our conclusion is negligible. Nonetheless we have adjusted all retrieval results with the additional broadening stemming from the observational setup. 

\subsection{Correction of stellar disk crossing effects}

During transit the planet obscures different regions of the stellar disk, which can introduce varying residual spectral lines at different phases. This contamination, due to the Rossiter-McLaughlin (RM) effect and limb-darkening, can be seen in 2D maps of the spectra as a so-called Doppler-shadow. Due to the geometry of the system and the planet velocity the RM-effect primarily impacts the phases around mid-transit for WASP-121~b and cannot impact the jet feature (see Figure \ref{fig:tracks}, where the sodium RM track has no overlap with the jet feature for the morning segment). We correct for this contamination in each individual in-transit epoch by modelling the stellar residual lines using \texttt{StarRotator}. \texttt{StarRotator} uses the rotational velocity $v\sin i$ to calculate the stellar spectrum locally obscured by the transiting planet as a function of orbital phase based on \citet{cegla_rossiter-mclaughlin_2016}. The needed stellar spectrum is generated via a \texttt{pySME} model \citep{wehrhahn_pysme_2023, Valenti1996} based on the VALD line list \citep{piskunov_vald_1995, ryabchikova_major_2015} with the stellar parameters from \citet{bourrier_hot_2020}. To compute the components of the spectra, the stellar surface is divided into $300$ x $300$ grid cells of different rotational velocities based on solid body rotation without differential rotation. We account for limb-darkening using the quadratic law \citep{kopal_detailed_1950}. For more information on the geometry and functionality of \texttt{StarRotator}, see \citet{prinoth_atlas_2024}.

 We have verified that no residuals beyond the noise-level from the RM-correction remain for our datasets by visually inspecting the transmission spectrum in the stellar rest frame (SRF) where any residual from the RM-correction would be most visible (see Figure \ref{fig:RM_residuals}). No such residual above the noise level was found in any of the master spectra around affected planetary lines.

\section{Parallel photometry}
\label{sec:photometry}

In parallel to the ingress epoch ESPRESSO observations, we observed the target with EulerCam \citep{lendl_wasp-42_2012} at the 1.2 meter Swiss Euler telescope at ESO's La Silla site. The images were taken in the Gunn-$r^{'}$ filter and the exposure time was set to 30 seconds. The raw full frame images were corrected for bias, over-scan and flat field, using the standard EulerCam reduction pipeline. We selected the optimal reference star and aperture combination via photometric scatter minimisation. The obtained light curve is shown in Extended Data Figure \ref{fig:photLightcurve}.
\noindent To constrain system parameters, we fitted the observed light curve with transit models using the code for transiting exoplanet analysis (CONAN, \citealp{lendl_signs_2017}). The jump parameters for the MCMC fit include mid-transit time, transit depth, impact parameter and (quadratic) limb-darkening coefficients.  In addition, we account for the correlated noise by iteratively fitting different baseline models involving time, air mass, FWHM and shifts of the stellar point spread function. The lowest Bayesian Information Criterion (BIC) is obtained for a baseline model involving a first order polynomial on FWHM of stellar point spread function. The constrained system parameters are reported in Table \ref{tab:parameters_W121}.
\noindent In addition, we searched for any occultation of active regions in the transit light curve by fitting a transit + spot model using PyTranSpot \citep{juvan_pytranspot_2018, chakraborty_sage_2024}. The jump parameters include mid-transit time, transit depth, latitude, longitude, temperature and size of active regions. The rest of the parameters were fixed to the values obtained from pure transit fitting. We found that a pure transit model has a significantly lower BIC than a transit + spot model. Thus, we conclude that our observations are not contaminated by any active region crossings.

\section{Determination of systemic velocity}
\label{sec:vsys_det}
We determined the systemic velocity by cross-correlating the out-of-transit exposures with a PHOENIX template of the star\citep{husser_new_2013}, following the methodology described in \citet{zhang_detecting_2023}. The systemic velocity is then determined by fitting a rotationally broadened model \citep{gray_observation_2008} including a polynomial of degree 4 for the continuum to the averaged out-of-transit cross-correlation function. The systemic velocity is $38.639 \pm 0.063$ \kms\ and $38.642 \pm 0.058$ \kms\ for ingress and egress respectively. The derivation of systemic velocities from spectroscopic data only includes the internal fitting errors and does not include systematic errors, e.g. from the lack of absolute wavelength calibrations due to the use of Fibre B for sky monitoring. However, the systemic velocity is a constant shift throughout the data and the difference between the two values measured on our data is below the resolution of one wavelength bin of ESPRESSO. The same is true for a comparison with previously recorded values in the literature on varying datasets \citep[$38.350\pm 0.021$ \kms][]{delrez_wasp-121_2016} and \citep[$38.36 \pm 0.43$ \kms][]{brown_gaia_2018} where again only the internal fitting error is accounted for leading to small uncertainties despite unknown accuracy.

\section{Atmospheric cross-correlation Doppler-trace}
\label{sec:planetaryCCF}
Based on the archival data of the egress epoch \citep[analysed in][]{borsa_atmospheric_2021,seidel_detection_2023-1}, and the described analysis of the newly obtained ingress epoch, we provide the phase-resolved spectra for four tracers in descending order: \halpha{}, \ion{Na}{I}, the full atmospheric cross-correlation, the \ion{Fe}{I} atmospheric cross-correlation. In the following we describe how the atmospheric cross-correlation traces are computed and compare our analysis to the literature results from \citet{borsa_atmospheric_2021}.\\
\noindent The planetary atmospheric track should follow along the prediction calculated from the planet's orbital parameters if no atmospheric movement is present. However, as observed e.g. in \citet{ehrenreich_nightside_2020}, the measured atmospheric track can deviate from the theoretical path thus tracking the phase-resolved line-of-sight velocity. The result of this analysis heavily depends on which spectral lines are compared to their theoretical position and the accuracy of said line lists. For example, this technique has successfully detected offsets in the planetary track for elements probing the deeper atmosphere in WASP-76~b \citep{ehrenreich_nightside_2020, kesseli_confirmation_2021} by employing a G2 stellar mask, thus mainly probing iron lines, and WASP-121~b \citep{borsa_atmospheric_2021}, using an F6 stellar mask under the assumption that stellar and planetary composition are comparable. Both planets have in consequence been the subject of various follow-up studies to fit global circulation models (GCMs) to said traces \citep[e.g. ][]{wardenier_modelling_2023,wardenier_phase-resolving_2024}. However, stellar masks, while accurate in line position, cannot give an indication of the altitude of the probed elements in the planetary atmosphere. We have therefore opted to use an optimised line list for planetary spectra at 2500~K from \citet{kitzmann_mantis_2023} with over $140$ species, including atomic species, individual ions, and H$_2$O, TiO, and CO for the full atmospheric trace. Additionally, from the same work, we have employed the pure \ion{Fe}{I} line list for an accurate iron trace compared to the approximation of stellar spectral masks. We create our cross-correlation analysis (comparison of the line list with the obtained spectra) of the here obtained dataset with {\tt tayph} \citep{hoeijmakers_hot_2020} and follow the atmospheric trace calculation from \citet{ehrenreich_nightside_2020}. An inspection of each exposure shows that nearly no planetary signal is present in the last in-transit exposure, indicating that a large fraction of the planet is already out of transit. This difference is most likely due to a small offset in transit timing compared to the earlier analysis in \citep{borsa_atmospheric_2021}. As a consequence, we have discarded the first and last in-transit exposures from our fits.\\
\noindent Visual comparison between our full atmospheric trace (reanalysing the existing egress epoch and combining with the new ingress epoch) and the full atmospheric trace from Figure 6. in \citep{borsa_atmospheric_2021} taken with ESPRESSO in 1-UT mode shows good agreement. We provide the average shift and FWHM for comparison purposes in Table \ref{tab:track_comparison}, where we have excluded true ingress and egress. Considering the five year gap between the observation of the two epochs, the 4-UT full atmospheric trace and the 1-UT full atmospheric trace, generated from an F6 mask, are in reasonable agreement, indicating a temporally stable atmosphere.\\

\noindent We provide the line-of-sight Doppler-shift for both the full atmospheric trace and the \ion{Fe}{I} trace in Figure \ref{fig:atmospheric_shimmer_all}. As many species are affected by the RM-effect, we have masked all phases between $-0.017$ and $0.017$. Considering our analysis of the Doppler-smearing, which adds an uncertainty of $1.1~\kms{}$ due to the movement of the planet during the exposure, the small Doppler-shift for the full trace leads to a nearly stationary deeper atmosphere for the morning segment. On the other hand, during the evening segment we observe a clear blue-shift indicating movement towards the observer. We highlight that the shifts provided in Table \ref{tab:track_comparison} are the line of sight velocity shifts as directly measured on the spectra and not the true velocity of the atmospheric dynamics. Taking into account the 3D nature of the planet and the viewing geometry, the measured line of sight velocity in the evening segment translates to an actual wind speed of $9.8\pm1.6\kms$. Separating the \ion{Fe}{I} trace shows a differing behaviour with overall lower velocities but a distinct blue-shift over both limbs, indicating a day-to-night side wind.\\

\section{\ion{Na}{I} doublet retrieval}
\label{sec:merc}

During transit observations, we can trace the movement of the atmosphere via the Doppler-shifts induced in the spectral lines as described for the atmospheric CCF trace. By focussing on single spectral lines instead of combining all lines from a species we gain the altitude information of these Doppler-shifts encoded in the line distortion. In the line distortion we additionally retain the information of asymmetrical winds (mostly via the line centre shift) and of symmetrical winds harder to track in cross-correlation via the additional broadening of the line. Combining this information as a function of altitude with a Bayesian retrieval framework can then discriminate between different atmospheric wind patterns at the origin of the line shape distortion. In short, instead of providing an estimate of the line-of-sight component of the global wind pattern, we obtain the dominant wind pattern with the true wind velocity. This resolved line wind retrieval algorithm is thoroughly documented in \citet{seidel_wind_2020,seidel_into_2021}. It applies pseudo-3D wind structures with planetary rotation and allows for two atmospheric layers (lower and upper) with distinct movement patterns for computational efficiency. The different atmospheric wind patterns are discriminated through a nested-sampling retrieval and the subsequent comparison of Bayesian evidence. 

For the archival egress epoch, this analysis was performed in \citet{seidel_detection_2023-1}. In this work, they explored the difference between the evening segment and the centre segment and found that a secondary blue-shifted feature only visible during the evening segment belongs to either an equatorially constrained day-to-night side flow or super-rotational jet. However, without the missing first transit third, these two scenarios are degenerate. They focussed their retrieval on the basic atmospheric wind patterns known from Solar System planets as well as the baseline case of no atmospheric dynamics: planetary rotation only with an isothermal temperature structure (iso), with additional super-rotational wind throughout the atmosphere (srot$_{\cos\theta}$), with additional day-to-night side wind throughout the atmosphere (dtn$_{\cos\theta}$). In the two cases of zonal winds the $\cos{\theta}$ dependence of solid body rotation is adopted to reduce the wind speed at the poles to zero. Additionally, they considered the onset of atmospheric mass loss with a symmetrical upwards, vertical wind (ver) as well as a basic combination of lower zonal winds and an upper atmospheric loss profile (two layer solutions). The best fit result to both obtain the broadening of the main \ion{Na}{I} feature, as well as the secondary, separated in wavelength space, blue-shifted peak was obtained by applying a zonal particle flow from the hot day-side to the cold-night side in the lower layer of the two layer model as well as a vertically outwards bound wind in the upper layer with a switch between layers roughly after one third of the probed pressure ranges. Additionally, Bayesian evidence revealed that the zonal flow had to be constrained to less than half the latitude range along the equator resulting in a localised jet stream across one limb. From the viewing geometry of the planet during the evening segment, this motion was only visible over the trailing limb and thus probed the evening terminator. Given the lack of data probing the other limb, they could thus not discriminate between a day-to-night side wind or a super-rotational jet stream.\\

For consistency with the detailed analysis of the evening segment as shown in \citet{seidel_detection_2023-1}, in this work we explore the same atmospheric wind profiles and adopt the latitude constraint of the jet stream profile of $\sim 20-40^\circ$ along the equator for the analysis of the morning segment and its prominent secondary \ion{Na}{I} feature - albeit now red-shifted compared to the blue shift for the evening segment. As customary in the Bayesian framework, we provide prior ranges informed by physics for all parameters. The prior ranges were selected as uniform priors with large ranges to capture potential degeneracies. The isothermal temperature ranges from $1500-5000$~K, the continuum degeneracy parameter NaX from $-4$ to $-1$. NaX encompases the degeneracy between the base pressure, the white light radius, and the abundance of sodium. We have opted throughout the manuscript to assume solar abundances for consistency. Given that a change in white light radius has an insignificant impact on the base pressure \citep{ Welbanks2019}, NaX directly adjust the pressure scale. All velocity ranges run from $0.1~\kms$ to $40~\kms$, which we apply as a generous upper limit for physically sensible velocities for atmospheres. The parameter space opened by these priors is then explored via Bayesian statistics and for each model the Bayesian evidence $|\ln\mathcal{Z}|$ is calculated together with a best fit and explored via a nested-sampling retrieval in the {\tt pymultinest} implementation \citep{feroz_2008, buchner_pymultinest_2016} and 6000 live points each. The best-fit parameters are the median of the marginalised posterior distributions. The advantage of this approach is the comparability of models via the difference in their Bayesian evidence $|\ln\mathcal{B}_{01}|$ when retrieved on the same dataset. The significance of this difference is judged via the Jeffrey's scale which can be found in \citet{trotta_bayes_2008}. \\

\noindent For our retrieval on the morning segment, we show the Bayesian evidence for each model in Extended Data Table \ref{table:comparison_model}. The corresponding posterior corner plots are presented in the Supplementary Material. We find a clear preference for a two layer model with a jet stream in the lower layer and outwards, vertical winds in the upper layer (see Extended Data Table \ref{table:comparison_model}). Additionally, we investigated whether the constraint of a jet is necessary to fit the data and find that, while the model with a fully super-rotational atmosphere in the lower layer provides a better fit than any of the single pattern models, the constrain to less than half the latitude range along the equator increases the Bayesian evidence by 1.5 orders of magnitude, in line with the findings on the evening segment in \citet{seidel_detection_2023-1}. An additional indicator for the continued presence of the jet-stream is the posterior distribution of both the model with a super-rotational wind in the entire atmosphere (see Supplementary Material). While the best fit with the highest probability for the fully super-rotational atmosphere is compatible with no wind, a secondary peak exists in the parameters space at the velocity of the jet-feature. For the two layer solution, while smeared significantly due to the unconstrained latitude of the lower layer winds, the velocity of the super-rotational wind also converges to the velocity of the jet-feature. On the other hand, the two layer model only converges tenuously on the velocity of the vertical particle movement in the upper layer. Most likely this is due to the low \ion{Na}{I} population at higher altitudes due to the relatively low retrieved temperature of $2000-2500$~K.

Putting the result into context with the analysis of the evening segment in \citet{seidel_detection_2023-1}, they found a jet-stream constrained to the same latitudes, however, at higher velocities than we find here during the morning segment (see Extended Data Table \ref{table:sodium_overview}). Additionally, the evening segment exhibits a temperature roughly $1000$~K above the morning segment. Due to this increased temperature, \ion{Na}{I} is found at higher altitudes and the vertical wind speed in the upper layer was better constrained for the evening segment. 
While we exclude the mid-transit exposures due to their possible contamination from stellar spectral line overlap (see Section \ref{sec:contaminants}), the centre segment was analysed in \citet{seidel_detection_2023-1} where they hypothesized that the zonal wind is obscured by night-side clouds during mid-transit as a consequence of the viewing geometry. Thus, only the upper-most layer of the atmosphere is probed by \ion{Na}{I}, which is corroborated by the increased temperature found for the centre third of the transit data. In support of this hypothesis and the lack of secondary \ion{Na}{I} features for the centre segment, models performed in \citep{helling_cloud_2021} for WASP-121~b suggest that the cloud cover at the anti-stellar point (night side) ranges from full coverage at 0.1 bar with high temperature condensates to approximately $40\%$ at $10^{-5}$ bar and beyond from metal oxide condensates. As a caveat, this analysis only extends to pressures of $10^{-5.5}$ bar and cloud coverage beyond this altitude remains speculative. Considering the probing ranges of the \ion{Na}{I}-traced jet stream beyond the modelled range, the night-side cloud hypothesis for the mid-transit phases remains plausible but unconfirmed. Considering the additional uncertainty from the impact of stellar lines on the line shape at these phases, we have opted to not include a discussion of the mid-transit data in the main body of the paper.

\section{\halpha{} retrieval}
\label{sec:pwinds}
The Balmer-series lines of hydrogen probe the upper atmospheric layers governed by the solar wind. The atmospheric structure at these altitudes is commonly described via Parker winds as first introduced in \citet{parker_dynamics_1958}. This atmospheric structure is implemented for Helium in the open-access {\tt p-winds} code \citep{dos_santos_search_2020}. We have implemented the Balmer-series in {\tt p-winds} by adopting the Boltzmann-Saha distribution and opacity calculations as described in \citet{wyttenbach_mass-loss_2020}. Most notably, we implement the Boltzmann equation for hydrogen for $\halpha$ following equation 7 in \citet{wyttenbach_mass-loss_2020} in local thermal equilibrium (LTE) as well as the non-LTE adjustment as described in equation 8 of the same work. The adjustment for non-LTE is specific to hydrogen in the thermosphere as probed by the Balmer-series lines, where the adjustment factor can be assumed constant \citep{huang_model_2017}. We additionally implement the canonical adjustment of the number of hydrogen atoms in a specific state from the Boltzmann equation via the Saha-equation for the hydrogen atom. The opacity is then calculated with the adjusted distribution of electron states via equation 9 of \citet{wyttenbach_mass-loss_2020}. Given that our analysis of the iron trace combined with the results from \citet{wardenier_phase-resolving_2024} favour an atmosphere dominated by hydrogen dissociation and recombination, we have opted to forgo the internal calculation of free electrons in {\tt p-winds} and have instead used the pre-computed electron density from model D of the fully hydrodynamical calculation of the atmosphere in \citet{huang_hydrodynamic_2023} which takes into account photoionsation, recombination and dissociation, but also charge exchange, collisional ionisation, radiative cooling, and the associated heating and ionisation from Lyman-$\alpha$ radiative transfer.\\ 
\noindent Due to the large absorption depth of the \halpha-line, we adopted a finer time-sampling than for the \ion{Na}{I} analysis. Supplementing the ingress epoch with the archival egress epoch, we bin the data by four exposures each, corresponding to the duration of ingress/egress. We retrieve the temperature ($T_0$), the mass loss rate ($\dot{m}$), and the central wavelength offset ($\delta \lambda$) via a {\tt pymultinest} implementation \citep{feroz_2008, buchner_pymultinest_2016} and 6000 live points each. From the retrieved values we calculate the horizontal movement from the central wavelength offset and the strength of the radial, outwards pointing Parker-winds. All results are provided in Extended Data Table \ref{tab:halpha_comparison} except for the mass loss rate, which could not be constrained, see Supplemental Material. Overall, the retrieval was unable to constrain the mass-loss rate $\dot{m}$ more than to the order of magnitude level from a prior spanning from $11^{-11}$ to $11^{-15}$. Considering that the hydrogen Balmer-series strength depends on the XUV flux of the star and is thus an ambiguous tracer for the outflow of the atmosphere, this result is unsurprising. Within this limitation, our retrieval provides mass-loss rates compatible with the literature. To calculate the best fit model shown in Figure \ref{fig:Halpha_resolved} we have opted to use the average escape value ($7 \cdot 10^{13} \frac{\mathrm{g}}{s}$) from the literature  \citep{huang_hydrodynamic_2023,koskinen_mass_2022}.

\section{Contribution in altitude}
\label{sec:contribution_func}

We computed the contribution function of the CCF for Fe, similar to the approach of \citet{molliere_petitradtrans_2019}, using cross-correlation templates. Each atmospheric layer $i$ in the template was iteratively set to zero opacity ($T_{\mathrm{ref}, i}$) and then compared to the nominal model, where opacity was retained in all layers ($T_{\mathrm{nom}}$), by calculating the contribution function as  $f_{\rm cont, i} = 1 - \frac{T_{\rm ref, i}}{T_{\rm nom}}$. These models were computed using \texttt{shone}, a radiative transfer code accounting for Rayleigh scattering and the H$^-$ opacity, assuming the same atmospheric structure and planetary parameters as \citet{kitzmann_mantis_2023} with solar abundance. To map this into velocity space, we computed the cross-correlation coefficients $C_i(v) = f_{\rm cont, i}T_{\rm ref, i}(v)$ between the nominal template and the contribution function $f_{\rm cont}$ and normalised each layer afterwards \citep{Pino2020}. Generally, cross-correlation templates are normalized to one, ensuring that \(\sum_{k=0}^N T_k(v) = 1\), where $k$ is the number of wavelength points/spectral pixel. However, we opted to turn off normalisation here to preserve the integrity of the contribution function to reflect the actual contribution of the layer. The resulting coefficients as a function of velocity $v$ are shown in Fig. \ref{fig:altitude_profile}. We additionally cross-validated the MERC contribution function via sodium.

We calculated the contribution function for transmission spectroscopy as described in \citet{molliere_petitradtrans_2019} for \ion{Na}{I} and \halpha{}. Each atmospheric layer in MERC and {\tt p-winds} has consecutively set its opacity to zero and the difference to the overall best-fit model is generated as described above for cross-correlation. This normalised difference then provides the contribution of each layer as a function of altitude to the model. The pressure scale is dynamically set taking into account scattering processes with constant solar abundance across all models \citep{ Welbanks2019}. Note that {\tt p-winds} provides a hydrodynamical calculation of the atmosphere as a function of radius, which can only be calculated for numerical reasons from 1.1 planetary radii. We then translated the radius profile to pressure via the ideal gas law as an approximation. To validate the altitude profile, we compared with the maximum probing depth estimation from \citet{borsa_atmospheric_2021} which arrived at the same maximum planetary radius for \halpha{}. 
We provide the altitude contribution profiles in Figure \ref{fig:altitude_profile}, with the best-fit model for \ion{Na}{I} of the morning segment, and the best-fit model from {\tt p-winds} for the first in transit exposure.

\clearpage

\end{methods}


\subsection{Data Availability}{Based on observations collected at the European Southern Observatory under ESO programmes 111.24J8 (PI: Seidel) and 1102.C-0744 (PI: Pepe, ESPRESSO GTO) at the European Southern Observatory. The data is publicly available on the ESO archive. All data products from the figures are available here: \url{https://github.com/jseideleso/vertical_jet_data}.}


\newpage




\begin{addendum}
\item The authors acknowledge the ESPRESSO project team for its effort and dedication in building the ESPRESSO instrument. This work relied on observations collected at the European Organisation for Astronomical Research in the Southern Hemisphere. We are especially grateful to Chenliang Huang for kindly providing us with the atmospheric structure (D) from \citet{huang_hydrodynamic_2023} and to Elspeth Lee for discussions on the limitations of the GCM in \citet{lee_mantis_2022}. The stellar modelling has relied on the PySME python package \citep{wehrhahn_pysme_2023}. All retrievals were performed with the {\tt pymultinest} package \citep{buchner_pymultinest_2016}. JVS gratefully acknowledges support from G. Chaparro and the Instituto de Física, Universidad de Antioquia. This research was supported by the Munich Institute for Astro-, Particle and BioPhysics (MIAPbP) which is funded by the Deutsche Forschungsgemeinschaft (DFG, German Research Foundation) under Germany´s Excellence Strategy – EXC-2094 – 390783311. This work was partially funded by the French National Research Agency (ANR) project EXOWINDS (ANR-23-CE31-0001-01). CFJ acknowledges support by the Swiss National Science Foundation (SNSF) under grant 215190. JPW acknowledges support from the Trottier Family Foundation via the Trottier Postdoctoral Fellowship. FPE would like to acknowledge the Swiss National Science Foundation (SNSF) for supporting research with ESPRESSO through the SNSF grants nr. 140649, 152721, 166227, 184618 and 215190. The ESPRESSO Instrument Project was partially funded through SNSF’s FLARE Programme for large infrastructures. J.L.-B. is funded by the Spanish Ministry of Science and Universities (MICIU/AEI/10.13039/501100011033/) and NextGenerationEU/PRTR grants PID2019-107061GB-C61 and CNS2023-144309. HMT acknowledges support from the “Tecnologías avanzadas para la exploración de universo y sus componentes" (PR47/21 TAU) project funded by Comunidad de Madrid, by the Recovery, Transformation and Resilience Plan from the Spanish State, and by NextGenerationEU from the European Union through the Recovery and Resilience Facility. Funded/Co-funded by the European Union (ERC, FIERCE, 101052347). Views and opinions expressed are however those of the author(s) only and do not necessarily reflect those of the European Union or the European Research Council. Neither the European Union nor the granting authority can be held responsible for them. This work was supported by FCT - Fundação para a Ciência e a Tecnologia through national funds and by FEDER through COMPETE2020 - Programa Operacional Competitividade e Internacionalização by these grants: UIDB/04434/2020; UIDP/04434/2020. A.S.M. acknowledges financial support from the Spanish Ministry of Science and Innovation (MICINN) project PID2020-117493GB-I00 and from the Government of the Canary Islands project ProID2020010129.  S.G.S acknowledges the support from FCT through Investigador FCT contract nr. CEECIND/00826/2018 and  POPH/FSE (EC). This work was supported by grants from eSSENCE (grant number eSSENCE@LU 9:3), the Swedish Research Council (project number 2023-05307$\_$VR), the Crafoord foundation and the Royal Physiographic Society of Lund, through The Fund of the Walter Gyllenberg Foundation. DE acknowledges funding from the Swiss National Science Foundation for project 200021$\_$200726 and from the National Centre of Competence in Research PlanetS supported by the Swiss National Science Foundation under grant 51NF40$\_$205606.
 
\subsection{Author contributions} 
JVS conceptualised the project as PI of the observing proposal, led the observational setup and execution of the program, led the analysis of the narrow-band features (\ion{Na}{I} doublet and $\halpha$), the retrieval on the \ion{Na}{I} doublet, as well as the p-winds update for $\halpha{}$ and the subsequent retrieval, and lastly led the manuscript elaboration and interpretation of all results.
BP created the atmospheric tracks and contribution function from cross-correlation for both the full atmosphere and iron.
LP contributed to the \ion{Na}{I} retrieval code and was instrumental in the creation of the quantitative vertical structure of the atmosphere.
LdS is the lead of the p-winds project and contributed to the extension of p-winds to the Balmer-series.
HC led the photometric analysis of the EulerCam data.
ES was instrumental in the data reduction and performed the telluric correction.
JPW provided the theoretical iron trails for WASP-121b and contributed to the interpretation of the observations.
CFJ was the observing astronomer for the supplementary EulerCam observations.
MRZ and RA provided input into the contaminants correction strategy.
DE contributed to the vertical profile interpretation.
VP contributed to the creation of the contribution functions and the comparison with GCMs.
ML was instrumental in scheduling the EulerCam observations.
GR contributed to the understanding of cloud formation and the subsequent implications for the results.
YD contributed to the atmospheric trace analysis and the overall data preparation.
FPE has been the PI of the ESPRESSO project and promoter of the 4-UT mode. In particular, he contributed in optimising the operations and the calibration, as well as the observations of WASP-121 in this mode.
SGS  contributed to the stellar characterisation.
HJH has contributed to the interpretation and contribution function.
All authors contributed in the preparation of the observing proposal and have read and commented the manuscript.

\subsection{Competing Interests} The authors declare no competing interests.
\subsection{Additional information} \,
\subsection{Supplementary information} Supplementary Figures are provided in a standalone supplementary material document.,
\subsection{Correspondence and requests for materials} should be addressed to the corresponding author (JVS). 
\subsection{Peer review information} \,
\subsection{Reprints and permissions information} is available at www.nature.com/reprints.
\end{addendum}

 



\clearpage

\begin{extended}
\setcounter{figure}{0}
\setcounter{table}{0}
\renewcommand{\figurename}{Extended Data Fig.}
\renewcommand{\tablename}{Extended Data Table}

\begin{table*}
\begin{threeparttable}
\caption{Summary of the movement change between morning and evening segment for \ion{Na}{I}, highlighting the impact of the crossing of the hot day-side of the planet. For the jet, the positive sign indicated the net blueshift and the negative sign the net redshift. The $1\sigma$ uncertainties of the posterior distribution is shown. For the evening segment as analysed in \citet{seidel_detection_2023-1}: While the uncertainty stemming from Doppler-smearing was accounted for in \citet{seidel_detection_2023-1}, the vertical wind profile was not adjusted for this additional factor. This leads to a small overestimate of v$_{\mathrm{ver}}$ (within the uncertainty). We have therefore corrected the values of v$_{\mathrm{ver}}$ presented in \citet{seidel_detection_2023-1} by $1.1~\kms$ in this work.}
\label{table:sodium_overview}
\centering
\begin{tabular}{l c c c c c}
\hline
\hline
Data & T$_{\mathrm{iso}}$ [K]  &  NaX &  v$_{\mathrm{jet}_{\cos\theta}}$ [\kms] & v$_{\mathrm{ver}}$ [\kms]     \\
\hline
morning segment &   $2404\pm662$  & $-2.41\pm0.40$   &  $13.7\pm6.1$ & $16.6\pm10.1$ & this work\\
evening segment &   $3350\pm470$  &  $-3.1\pm0.20$  &  $-26.8\pm7.3$ & $26.7\pm9.3$ & \citet{seidel_detection_2023-1}\\
\hline
\end{tabular}
\end{threeparttable}
\end{table*}

   \begin{figure}[h!]
   \centering
\resizebox{\columnwidth}{!}{\includegraphics[trim=2.0cm 0.0cm 1.0cm 0.0cm]{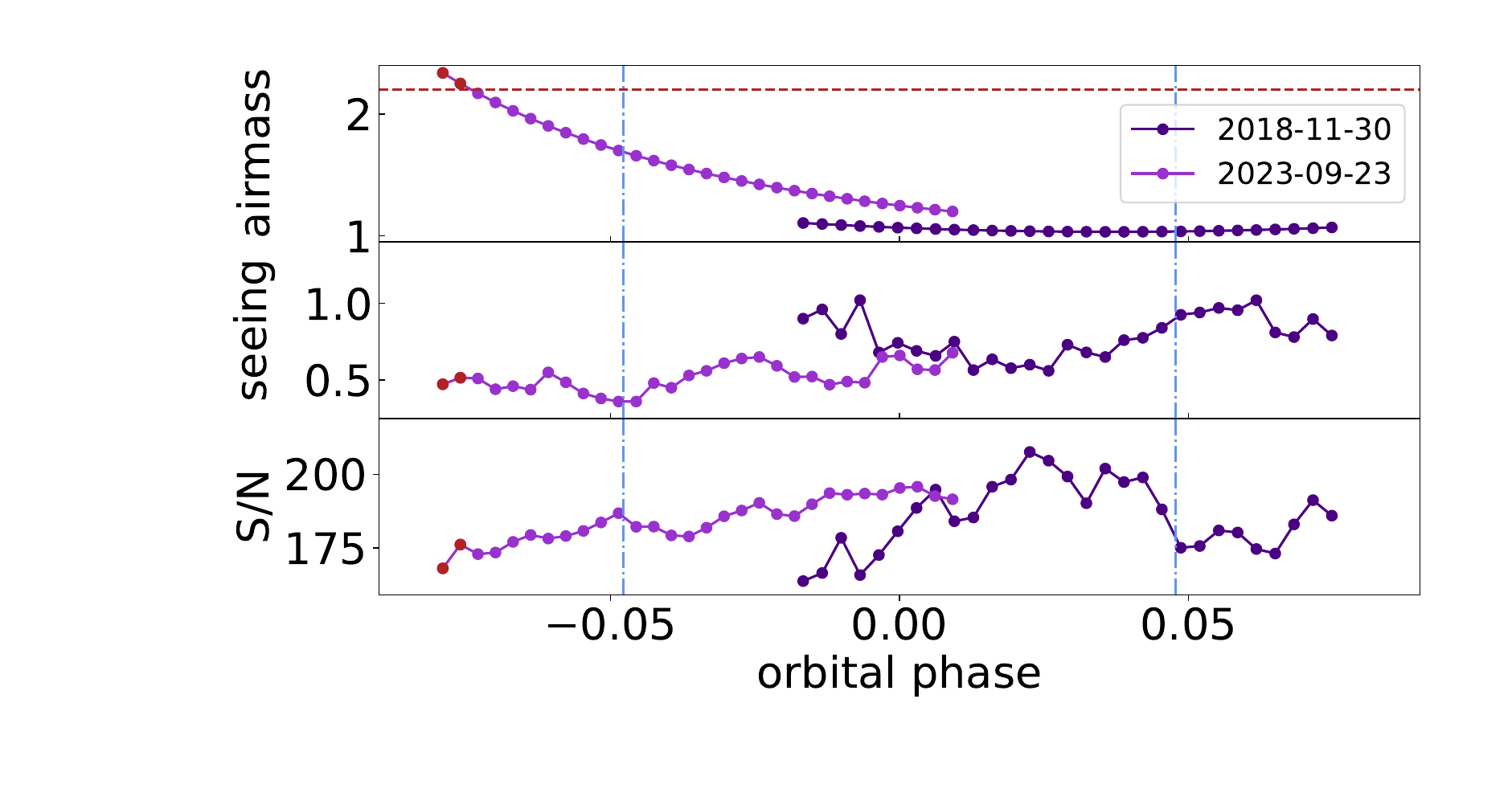}}
      \caption{Good conditions for both observations. The conditions are shown for the two partial transits, showing airmass, seeing, and signal to noise ratio as a function of phase. Ingress and egress are shown as vertical, blue, dashed-dotted lines, the ADC limit above which the data is not properly corrected for atmospheric dispersion is shown as a horizontal, red, dashed line in airmass. The two discarded exposures are marked in red above the ADC limit.}
         \label{fig:transitOverview}
   \end{figure}

   \begin{figure}
   \centering
\resizebox{\columnwidth}{!}{\includegraphics[trim=0.0cm 0.0cm 0.0cm 0.0cm]{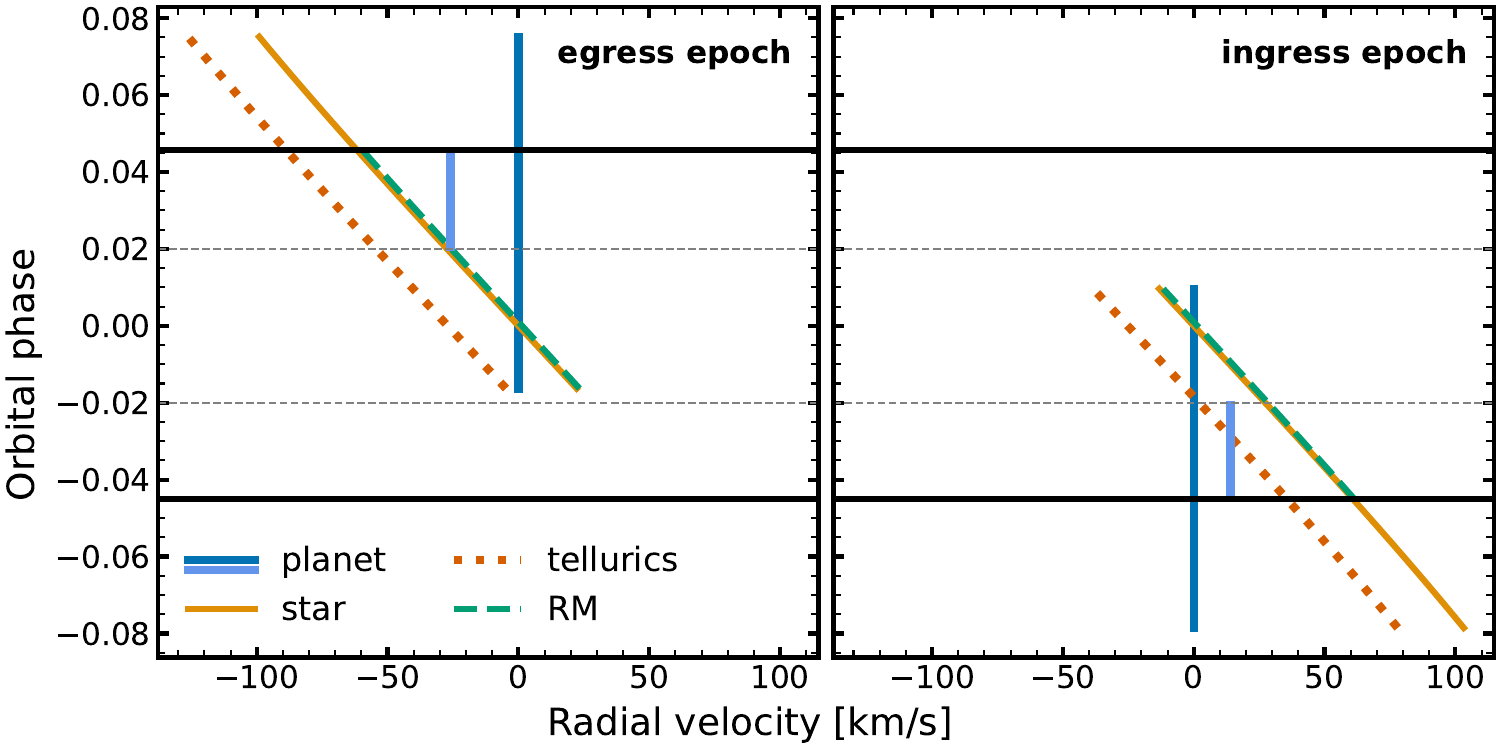}}
      \caption{Impact of contamination sources. The planetary trace for the observed phases (blue) and the trace of the jet (light blue) is shown for both epochs with the most important contamination source (RM, green dashed, stellar lines, golden solid, and tellurics, red dotted) in the planetary rest frame. The thick black lines mark the start and end of the transit, while the dotted horizontal lines mark the split of the data in morning, centre, and evening segment. It is immediately clear from this overview of the velocities that the RM-effect contaminates the centre third, which we have subsequently not taken into account. Additionally, telluric contamination has to be accounted for during the ingress third to claim the jet detection.}
         \label{fig:tracks}
   \end{figure}

\begin{figure}
   \centering
\resizebox{\columnwidth}{!}{\includegraphics[trim=5.0cm 2.5cm 1.0cm 1.0cm]{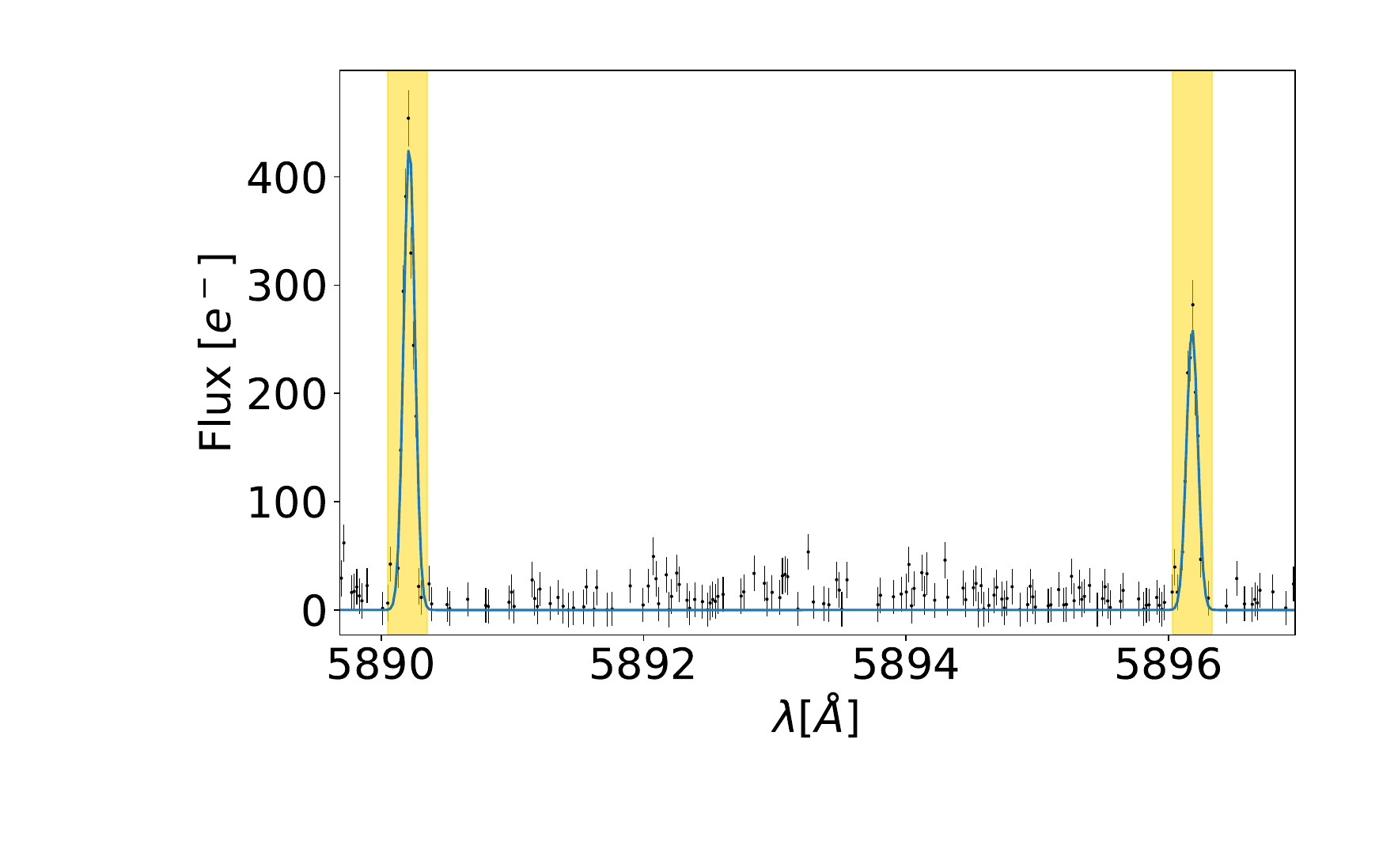}}
      \caption{Sky contamination from meteor showers. Fibre B spectrum (exposure 11) is shown in the observer's rest frame with a double Guassian fit to the \ion{Na}{I} emission peaks (in blue). The masked wavelength regions are shown as a yellow box. The $1\sigma$ uncertainties have been propagated accordingly from the errors calculated by the ESPRESSO pipeline. Given that the bias is substracted from the sky fibre with no additional background, some continuum values would be negative. These values are automatically removed by the pipeline leading to gaps in the data coverage.}
         \label{fig:sodium_emission}
   \end{figure}

   \begin{figure*}
   \centering
\resizebox{\textwidth}{!}{\includegraphics[trim=0.0cm 0.0cm 0.0cm 0.0cm]{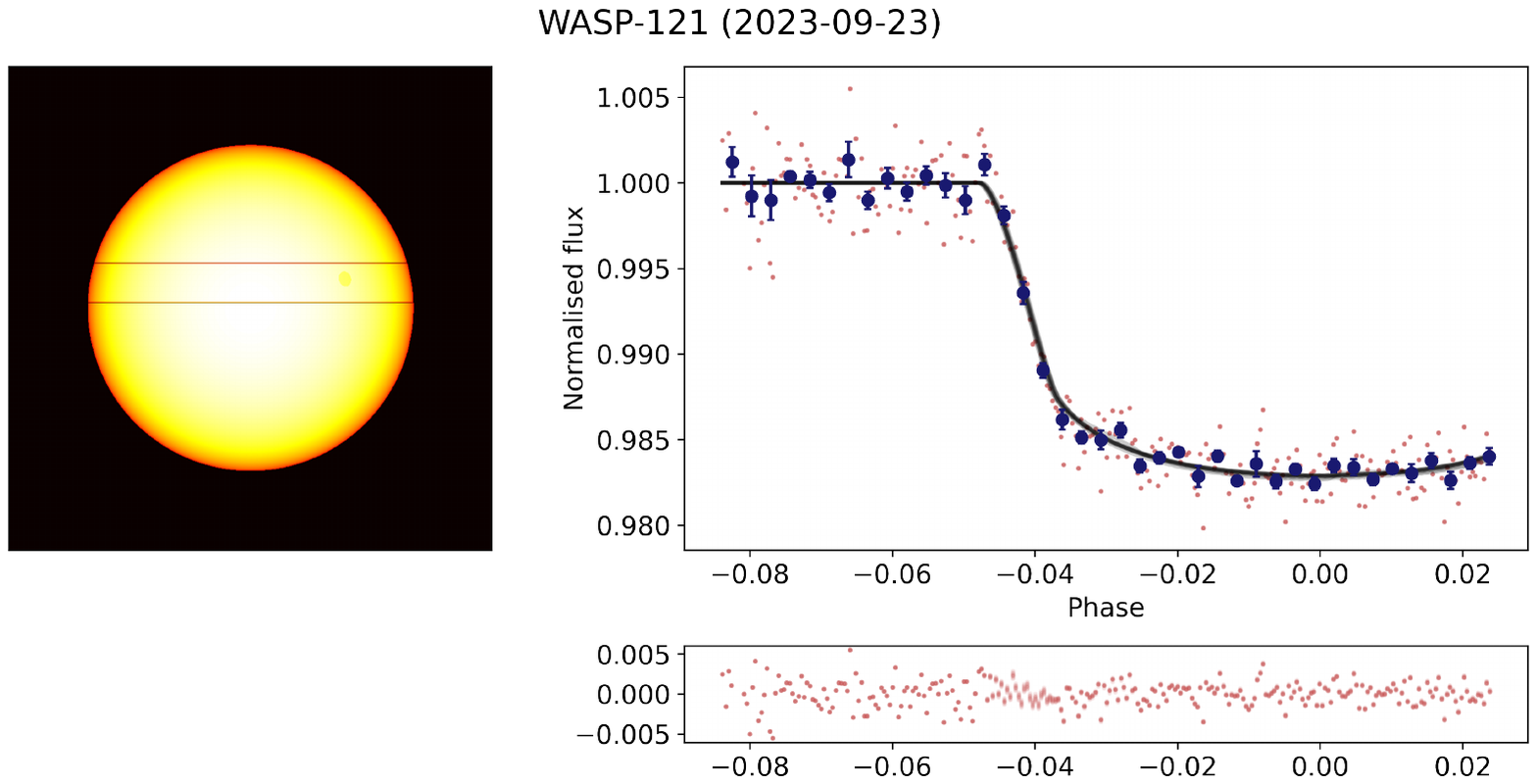}}
      \caption{Low stellar activity estimate from photometry observations. Left: {\tt PySpot} simulation of WASP-121 with the transit cord shown with two horizontal lines. Note that this plot only displays the lack of spots, the real orbit of WASP-121~b is close to polar. This geometry feature is unfortunately not available in {\tt PySpot}. Right: EulerCam lightcurve observations (binned by 5 in black) with the best fit {\tt conan} model as a black, solid line with the residuals of the fit below. The $1\sigma$ uncertainties have been propagated accordingly from the errors calculated by the EulerCam pipeline.}
         \label{fig:photLightcurve}
   \end{figure*}

\begin{figure}
   \centering
\resizebox{\columnwidth}{!}{\includegraphics[trim=3.0cm 1.8cm 0.0cm 1.0cm]{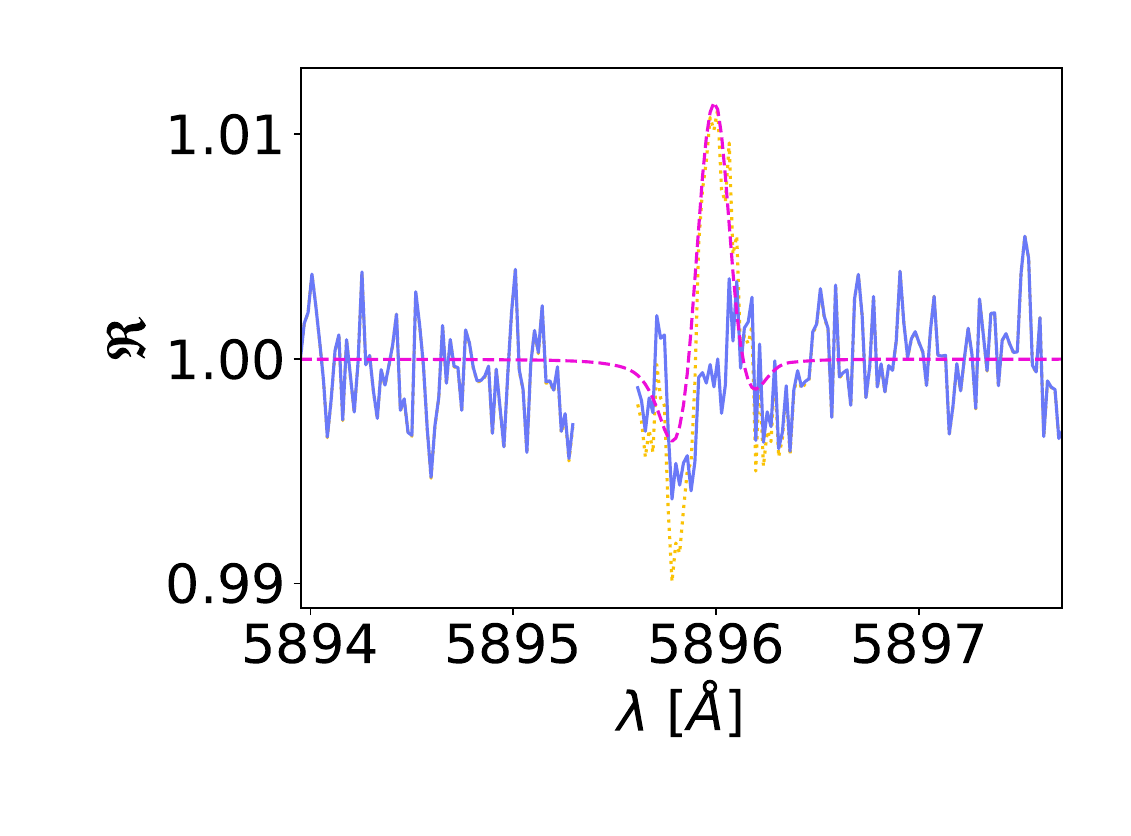}}
      \caption{Assessment of the negligible impact of the RM-effect. The RM-effect correction around the \ion{Na}{I} D$_1$ Fraunhofer line in the stellar rest frame (SRF) is shown for the in-transit master spectrum divided by the out-of-transit master spectrum, the so called SRF transmission spectrum. In golden the RM-uncorrected data is shown, with the mean of all RM-models for all in-transit spectra in fuchsia as an overlay. The corrected SRF transmission spectrum is shown in purple. The data discontinuity arises from the masking of the telluric emission feature in the observer's rest frame. The $1\sigma$ uncertainties have been propagated accordingly from the errors calculated by the ESPRESSO pipeline.}
         \label{fig:RM_residuals}
   \end{figure}

\begin{table}
        \caption[]{Summary of the stellar and planetary parameters of the \targetsys system adopted in this study. References: [1] \cite{delrez_wasp-121_2016}, [2] \cite{borsa_atmospheric_2021}, [3] \cite{bourrier_hot_2020}, [*] this work}
        \label{tab:parameters_W121}
        \small
        \begin{center}
                \def\arraystretch{1.25}
                \begin{tabular}{p{0.52\linewidth}p{0.33\linewidth}p{0.025\linewidth}}
                        \toprule
            \multicolumn{3}{c}{\targetsys system parameters}  \\ \midrule
            RA$_{\rm 2000}$ & 07:10:24.06 & [1]\\ 
                DEC$_{\rm 2000}$ & -39:05:50.55 & [1]\\ 
                Distance [pc] & \num{270(90)} & [1] \\
                Magnitude [V$_{\rm mag}$] & 10.4 & [1]\\
                Sys. velocity ($v_{\rm sys}$) [\si{\km\per\second}]    & \num{38.198(0.002)} & [2] \\ 
                 \midrule
                \multicolumn{3}{c}{Stellar parameters} \\ \midrule
                Star radius ($R_\ast$) [$R_{\odot}$]    & \num{1.44(0.03)} & [2]\\
                        Star mass ($M_\ast$) [$M_{\odot}$] & \num{1.38(0.02)} & [2] \\
                        Proj. rot. velocity ($v\sin{i}$) [\si{\km\per\second}] & \num{11.8(0.2)} & [2] \\
                Age [Gyr] & \num{1.03(0.43)} & [2] \\
                Metallicity [Fe/H] & \num{0.13(0.04)} & [2] \\
                        \midrule
                        \multicolumn{3}{c}{Planetary parameters}  \\ \midrule
                        Planet radius ($R_{\rm p}) $ [$R_{\rm Jup}$]  & \num{1.7402(0.0393)} & [*]    \\
                        Planet mass ($M_{\rm p} $) [$M_{\rm Jup}$]  & \num{1.157(0.070)} & [3] \\
                        Eq. temperature ($T_{\rm eq}$) [$\si{\kelvin}$] & \num{2358(52)} & [1] \\
                        limb darkening u1 & \num{0.300(0.088)} & [*] \\
                        limb darkening u2 & \num{0.065(0.191)} & [*] \\
                        \midrule
                        \multicolumn{3}{c}{Orbital and transit parameters} \\ \midrule
                        $T_0$ [HJD (UTC)]  & \num{2460211.8724(0.0004)} & [*] \\
                        Semi-major axis ($a$) [au]   & \num{0.02582(0.00055)} & [*] \\
                        Scaled $a$ ($a/R_\ast$)  & \num{3.81(0.02)}& [*]\\
                        Orbital incl. ($i$) [$^\circ$]   & \num{87.7(0.6)} & [*]  \\
                        Proj. orbital obliquity ($\lambda$) [$^\circ$] & \num{-87.08(0.28)} & [2] \\
                        Eclipse duration ($T_{14}$) [h] & \num{2.9053(0.0060)} & [3]\\
                        Radius ratio ($R_p/R_\ast$) & \num{0.1227(0.0011)} & [*] \\
                        RV semi-amp. ($K$)  [\si{\km\per\second}] & \num{177(8.5)} & [3]\\
                        Period ($P$) [d]   & $1.27492504\pm 1.5E-7$ & [3]  \\
                        Eccentricity & $0$ & [1]  \\
                        Impact param. ($b$) & \num{0.10(0.01)}  & [3]  \\
                        $v_{\rm orb}$  [\si{\km\per\second}]    & \num{137.8(2.2)} & [*]\\   
                        \bottomrule
                \end{tabular}
        \end{center}
 \end{table}

 \begin{table*}
    \centering
    \caption{Line-of-sight fitting values of the atmospheric tracks. The $1\sigma$ uncertainties have been propagated accordingly from the errors calculated by the ESPRESSO pipeline. The planet template (planet) where we generate the mask for all likely species present in the planet from \cite{kitzmann_mantis_2023}, including iron. \cite{borsa_atmospheric_2021}  applies an F6 stellar type template as an approximation of the likely planetary composition (F6 stellar).}
    \label{tab:track_comparison}
    \begin{tabular}{lccccl}
&  shift morning & shift evening & width morning & width evening  & (all in $\kms$) \\ \hline
\ion{Fe}{I} 4-UT & $-4.12\pm0.15$ & $-6.90\pm0.11$ &  $15.0\pm0.4$ &  $13.1\pm0.3$ & this work \\
full 4-UT (planet) & $-2.68\pm0.14$ & $-7.51\pm0.16$ &  $11.9\pm0.3$ &  $10.7\pm0.3$ & this work \\
full 1-UT (F6 stellar) & $-2.80\pm0.28$ & $-7.66\pm0.16$ &  $7.3\pm0.7$ &  $10.2\pm0.4$ & \citet{borsa_atmospheric_2021} \\ \hline
    \end{tabular}
\end{table*}

\begin{table}
\begin{threeparttable}
\caption{Comparison of the different models for the morning segment. The $1\sigma$ uncertainties of the posterior distribution are shown. The base model to calculate $|\ln\mathcal{B}_{01}|$ is the isothermal model with no added wind patterns (none). The model with the highest Bayesian evidence is highlighted in bold.}
\label{table:comparison_model}
\centering
\begin{tabular}{l c c c c c}
\hline
\hline
Model & $|\ln\mathcal{Z}|$   &  $|\ln\mathcal{B}_{01}|$ & Odds & Probability ($\sigma$) & p-value  \\
\hline
none  &  $384.60\pm0.20$   & -  & - & - & - \\ 
$\mathrm{srot}_{\cos\theta}$  &  $382.75\pm0.14$   & $-1.85$  & $0.2:1$ & $13.6\%(-1.1)$  & $0.864$ \\
$\mathrm{dtn}_{\cos\theta}$  &  $383.74\pm0.15$   & $-0.86$  & $0.4:1$ & $29.6\%(-0.5)$  & $0.704$ \\
ver  &  $384.87\pm0.25$   & $0.27$  & $1.3:1$ & $56.5\%(1.0)$ & $0.053$\\
$\mathrm{srot}_{\cos\theta}, \mathrm{ver}$  &  $387.11\pm0.10$   & $2.51$  & $13:1$ & $92.6\%(2.7)$ & $0.006$\\
$\mathrm{\textbf{jet}}_{\cos\theta}, \mathrm{\textbf{ver}}$  &  $388.62\pm0.26$   & $4.02$  & $56:1$ & $98.2\%(3.3)$ & $0.001$\\
\hline
\end{tabular}
\end{threeparttable}
\end{table}

\begin{table*}
\begin{threeparttable}
    \centering
    \caption{Fitting values of the \halpha{} line. The $1\sigma$ uncertainties of the posterior distribution is shown. The velocity offset of the line centre ($\delta v$) adjusted for the line of sight component of the planetary rotation for the first and last set which is true ingress and egress. The planetary rotation is calculated under the assumption of synchronous rotation with the star. Additionally, all velocities translated from the line of sight velocity to the actual wind speed taking into account the 3D nature of the planet and the viewing geometry.}
    \label{tab:halpha_comparison}
    \begin{tabular}{cccc}
phase &  T [K]  & $\delta \lambda$  & $\delta v_{\mathrm{adj}} [\frac{km}{s}]$\tnote{a} \\
\hline

$-0.041$ & $9360\pm280$ &  $0.12\pm0.03$ & $5.2\pm1.4$\\
Ingress end &  & & \\ 
$-0.029$ & $8700\pm260$&  $0.03\pm0.02$&  $2.3\pm0.9$\\
$-0.017$ & $8500\pm260$&  $-0.04\pm0.02$ & $-2.9\pm0.9$\\
Overlap with the RM-effect & & & \\
$0.021$ & $8850\pm270$&  $-0.12\pm0.02$ & $-8.9\pm0.9$\\
$0.034$ & $8950\pm270$&   $-0.14\pm0.02$& $-10.3\pm0.9$\\
Egress start &  & & \\
$0.044$ & $9800\pm340$&  $-0.31\pm0.05$& $-19.2\pm1.4$\\
 
\hline
    \end{tabular}
\end{threeparttable}
\end{table*}

\end{extended}

\end{document}